\documentclass[aps,prb,superscriptaddress,groupedaddress,floatfix,twocolumn,showpacs]{revtex4}
\usepackage{graphicx}
\usepackage{dcolumn}
\usepackage{bm}
\usepackage{bbold}
\usepackage{stmaryrd}
\usepackage{latexsym}
\usepackage{amssymb}
\usepackage{amsfonts}
\usepackage{amsmath}
\usepackage{mathtools}
\usepackage{fancybox}
\usepackage{color}
\usepackage[breaklinks=true,colorlinks,citecolor=blue,linkcolor=blue,urlcolor=blue]{hyperref}

\begin{document}
\title{Orbital and spin order in oxide two-dimensional electron gases}
\author{John R. Tolsma}
\email{tolsma@physics.utexas.edu}
\affiliation{Department of Physics, The University of Texas at Austin, Austin Texas 78712, USA}
\author{Marco Polini}
\affiliation{Istituto Italiano di Tecnologia, Graphene Labs, Via Morego 30, I-16163 Genova,~Italy}
\author{Allan H. MacDonald}
\affiliation{Department of Physics, The University of Texas at Austin, Austin Texas 78712, USA}
\date{\today}
\begin{abstract}
We describe a variational theory of multi-band two-dimensional electron gases that captures the interplay between electrostatic confining potentials, orbital-dependent interlayer electronic hopping and electron-electron interactions, and 
apply it to the d-band two-dimensional electron gases that form near perovskite oxide surfaces and heterojunctions.  
These multi-band two-dimensional electron gases are prone to the formation of Coulomb-interaction-driven orbitally-ordered 
nematic ground-states.  We find that as the electron density is lowered and interaction effects strengthen,  
spontaneous orbital order occurs first, followed by spin order. 
We compare our results with known properties of single-component two-dimensional electron gas systems and comment on closely related physics in semiconductor quantum wells and van der Waals 
heterostructures.  
\end{abstract}
\maketitle

\section{Introduction}
\label{sect:Intro}
Orbital and spin order are relatively ubiquitous in strongly interacting systems, and have long been studied in ${\rm ABO_3}$ bulk crystals with the perovskite crystal structure.~\cite{wollan_pr_1955,goodenough_pr_1955}  In perovskites 
B is typically a transition metal and A is either an alkaline or rare earth metal. Among these materials, those with partially filled transition metal $t_{2g}$ d-orbitals are particularly interesting because orbital and spin order appear almost simultaneously near the magnetic transition temperature.~\cite{ren_prb_2009,yan_prl_2004,cheng_prl_2008}  In crystals for which the number of $t_{2g}$ electrons per transition-metal is close to one, the probability of two electrons occupying the same site is substantial, and short-range Hubbard-type interaction models capture the most important parts of the interaction physics.~\cite{mizokawa_prb_1996,mochizuki_jpsj_2004}  Generalizations of the Hubbard model apply for larger integer values of the number of electrons per metal.  
As a consequence of relatively recent advances~\cite{stemmer_apl_2013,hwang_apl_2010,mannhart_science_1010,stemmer_armr_2014,levy_armr_2014} it is now possible to add electrons to quantum wells formed by $d^{0}$ perovskites, most commonly SrTiO$_3$.
The resulting systems are two-dimensional metals with far fewer than one conduction electron 
per transition-metal site. Because of its long range, the typical Coulomb
interaction energy of an individual electron in these systems drops to zero only as two-dimensional density $n^{1/2}$, 
in contrast to the $\propto n$ behavior of the Hubbard model.
Therefore, in these systems the full long-range Coulomb interaction must be retained~\cite{Tolsma_PRB_2016} in order to 
achieve a realistic description of electron-electron interaction effects. In this Article we address orbital and spin order,
and their interplay in low carrier-density perovskite quantum well systems.
 
The possibility of spin order in electron gases with a single parabolic band was raised very early in the 
history of the quantum theory of solids.~\cite{Bloch_1929}
Although many experiments have hinted at ferromagnetic instabilities in these single-band electron gas
systems,~\cite{young_nature_1999,vitkalov_prl_2001,shashkin_prl_2001,teneh_prl_2012,takashina_prb_2013,renard_naturecommun_2015}
unambiguous evidence for magnetism has so far been lacking. 
Theoretically, order is expected only at extremely low carrier densities, at 
least in the absence of disorder.~\cite{dassarma_prb_2005,attaccalite_prl_2002,tanatar_prb_1989}
Our study captures the peculiarities of perovskite multi-band two-dimensional electron gas systems 
which open up a richer range of experimental possibilities.  
\begin{figure}[h!]
\includegraphics[width=0.95\linewidth]{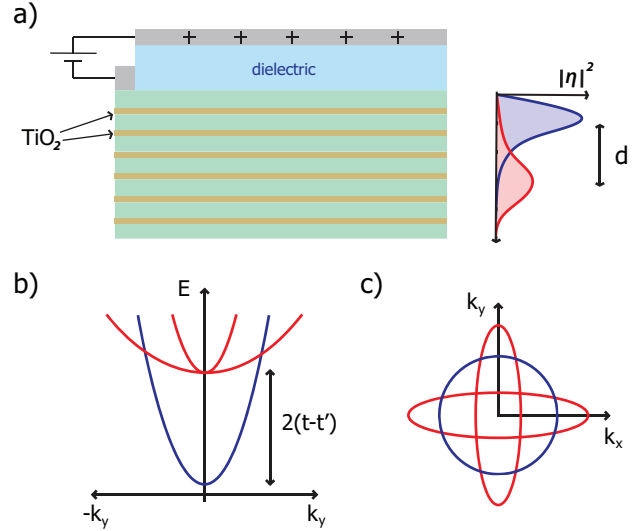}
\caption{(Color online)  A schematic summary of the $t_{2g}$ 2DEG model for SrTiO$_3$. 
a) At left is a top-gated SrTiO$_3$ heterojunction which hosts $t_{2g}$ electrons distributed across two-dimensional
TiO$_2$  layers highlighted in orange. 
At right is a sketch of the squared amplitude of the subband wavefunctions, $\vert{\bm \eta}\vert^2$, for the $t_{2g}$ 2DEG model described in Section~\ref{Sect:One}. The $xz$ and $yz$ subband wavefunctions are shown in red and the $xy$ subband wavefunction in blue, illustrating the difference between their interfacial confinements. The energy offset, equal to $2(t-t^{\prime})$ in the single-layer limit,
between the anisotropic $xz$ and $yz$ (red) band edges and the isotropic $xy$ (blue) band edge is illustrated in b) 
and the anisotropy of the elliptical $xz$ and $yz$ Fermi surfaces (red) is contrasted with the $xy$ band's circular Fermi surface (blue)
in c).  $t$ and $t$' are respectively the large and small hopping amplitudes for $t_{2g}$ orbitals discussed 
in the main text. \label{fig:One}}
\end{figure}

In multiband two-dimensional electron gas heterojunctions and quantum wells, the ground-state energy
depends sensitively on the density in each spin-resolved band. 
We propose a modified variational theory which captures the interplay between orbital-dependent interlayer electronic hopping, the electrostatic confining potential, and electron-electron interactions.
Our theory attempts to capture physics that is missing in most 
density-functional-theory~\cite{Hohenberg_Kohn_PhysRev_1964,Kohn_Sham_PhysRev_1965} calculations 
which typically employ extended version of the local-density approximation (LDA) for the exchange-correlation (XC) energy functional, 

\begin{equation}\label{eq:LDA}
E^{\rm LDA}_{\rm XC}\big[n\left({\bm r}\right)\! \big] = \int \! d{\bm r}\, n\!\left({\bm r} \right)\, \xi_{\rm XC}\big(n\!\left({\bm r}\right)\! \big)~, 
\end{equation}
where $\xi_{\rm XC}\big(n\!\left({\bm r}\right)\! \big)$ is the XC energy-per-electron of the single-band isotropic electron gas model with homogeneously distributed electron density $n\!\left({\bm r} \right)$. Despite its great success, it is generally known that the LDA and its 
generalizations, ({\it e.g.} the generalized gradient 
approximation~\cite{Langreth_PRB_1980,Perdew_PRL_1996}) do not in all cases adequately describe the XC energy. 
In covalent semiconductors, this shortcoming is highlighted by a significant underestimation of the band-gap energy. 
This limitation can be traced back to Eq.~(\ref{eq:LDA}), where we see that the LDA XC energy functional is sensitive to the local 
electron density but not to the orbital symmetries of the occupied electronic states~\cite{Giuliani_and_Vignale,Jones_RMP_1989}. 
Motivated by an interest in orbital and spin order in $t_{2g}$ two-dimensional electron gases, we have developed a broadly 
applicable variational theory for the ground-state energy which is sensitive to the electron density in each band, and therefore is 
sensitive to electron orbital symmetry in systems in which each band has a dominant orbital character.

Our article is organized as follows. In Section~\ref{Sect:One} we briefly summarize the $t_{2g}$ 2DEG model we have 
introduced previously,~\cite{Tolsma_PRB_2016} and extend it to account explicitly for the concerted action of orbital-dependent electronic hopping along the confinement direction and electron-electron interactions. 
In Section~\ref{Sect:Two} we describe a variational approximation for the ground-state that accounts self-consistently for interfacial confinement forces and their influence on XC energies.  
As a first application of this method, we calculate the phase diagram for the $t_{2g}$ 2DEG as a function of 
total density and quantum well thickness in Section~\ref{Sect:Three}. 
We show that Coulomb interactions give rise to a novel sequence of orbital- and spin-ordered ground-states, with correlations 
favoring the former. 
Compared to the single-component isotropic 2DEG case, the $t_{2g}$ 2DEG is
dramatically more susceptible to these 
Bloch-like~\cite{Bloch_1929} instabilities. We comment on the experimental signatures of these states and 
suggest possible extensions of the calculations presented here. 
Finally, in Section~\ref{Sect:Four} we summarize our results and comment on the applicability of our variational approximation
to semiconductor quantum wells, van der Waals heterostructures, and other multi-component 2D electron systems.

\section{$t_{2g}$ electron gas model}
\label{Sect:One}
We begin by summarizing the simple model for the low-energy conduction bands in ${\rm SrTiO}_3$ quantum wells and 
heterojunctions that we study. 
We first separate the non-interacting part of the
Hamiltonian into two sets of terms: those which describe electronic motion parallel ($\parallel$) to the 
confinement direction, and those which describe electronic motion perpendicular ($\perp$) 
to the confinement direction, which we take to be the $\hat{\bm z}$ direction. 
The purpose of this separation will become clear in Section~\ref{Sect:Two}. 

The band structure of ${\rm SrTiO}_3$ electron gases has been thoroughly discussed in the 
literature.~\cite{guru_prb_2012,millis_prb_2013,Popovic_prl_2008}
Here we briefly describe some details required for the present analysis.
The low-energy conduction bands of SrTiO$_3$ quantum wells and heterojunctions are 
formed from the $xy$, $yz$, and $xz$ d-orbitals (~the $t_{2g}$ orbitals) of Ti. SrTiO$_3$ has a perovskite crystal structure, but the sublattice formed by the Ti atoms is simple cubic and described by the lattice vector $\bm{R} = a\left(n_{x}\hat{\bm x} + n_{y}\hat{\bm y} +n_{z}\hat{\bm z}\right)$ where $a = 3.9$~\AA~is the lattice constant of SrTiO$_3$. Orbital symmetry dictates that each $t_{2g}$ electron hops mainly between states with the same orbital character, with a large hopping amplitude $t$ in the two directions in the plane of the 
orbital, and a smaller hopping amplitude $t^{\prime} $ in the third direction.~\cite{guru_prb_2012}
The weak hopping directions for $xy$, $xz$, and $yz$ orbitals are $\hat{\bm z}$, $\hat{\bm y}$, and $\hat{\bm x}$, respectively. Nearest neighbor hopping on the simple-cubic Ti sublattice is described by the non-interacting Hamiltonian
\begin{equation}
{\cal H}_{0} = \displaystyle{\sum_{\alpha \sigma}}\displaystyle{\sum_{\bm{R}\bm{R}^{\prime}}}  c^{\dagger}_{\bm{R} \alpha \sigma} \, h_{\alpha}(\bm{R}-\bm{R}^{\prime}) \, c_{\bm{R}^{\prime} \alpha \sigma}~,
\end{equation}  
where 
\begin{widetext}
\begin{equation}\label{eq:hopping}
h_{\alpha}(\bm{R}-\bm{R}^{\prime}) = \left\{
\begin{array}{l}
{\displaystyle   -t \, \delta_{\Delta\bm{R},\pm a \hat{\bm x}} -t \, \delta_{\Delta\bm{R},\pm a \hat{\bm y}} -t^{\prime} \, \delta_{\Delta\bm{R},\pm a \hat{\bm z}} \qquad \,\, {\rm for} \,\, \qquad \alpha=xy}\vspace{0.3 cm}\\
{\displaystyle -t\, \delta_{\Delta\bm{R},\pm a \hat{\bm x}} -t^{\prime}\, \delta_{\Delta\bm{R},\pm a \hat{\bm y}} -t\, \delta_{\Delta\bm{R},\pm a \hat{\bm z}}  \,\, \qquad {\rm for} \qquad \,\, \alpha=xz}\vspace{0.3 cm}\\
{\displaystyle -t^{\prime}\, \delta_{\Delta\bm{R},\pm a \hat{\bm x}} -t \, \delta_{\Delta\bm{R},\pm a \hat{\bm y}} -t \, \delta_{\Delta\bm{R},\pm a \hat{\bm z}} \,\, \qquad {\rm for} \,\, \qquad  \alpha=yz}~,
\end{array}
\right.~
\end{equation}
\end{widetext}
where the electron spin is labelled by $\sigma$ and we have defined $\Delta\bm{R} = \bm{R}-\bm{R}^{\prime} $. 

2DEGs have been realized in $\delta$-doped SrTiO$_3$,~\cite{hwang_apl_2010} in ionic-liquid 
gated SrTiO$_3$,~\cite{Ueno_natmater_2008} at heterojunctions between SrTiO$_{3}$ and polar 
perovskites,~\cite{ilani_natcomm_2014,stemmer} and
in electrostatically gated SrTiO$_3$.~\cite{Goldman_PRL_2011}
For definiteness, we consider a SrTiO$_3$ heterojunction extending from $z=0$ to $z=a(N_l+1)$ 
with an electrical gate placed at 
$z <0$. Here, $N_l$ is the number of TiO$_2$ layers (see Figure~\ref{fig:One}) with appreciable occupation.
Electron motion in the plane perpendicular to the $\hat{\bm z}$-direction is unbounded and Bloch bands are formed from $t_{2g}$-orbital hopping on the square Ti 2D sublattices. 
Electron creation operators in real and 2D momentum space are related by
\begin{equation}
c^{\dagger}_{\bm{R} \alpha \sigma} = \frac{1}{\sqrt{N}}\displaystyle{\sum_{\bm{k}}}e^{-i\bm{k}\cdot\bm{R}_{\perp}}c^{\dagger}_{\bm{k} R_z\alpha \sigma},
\end{equation}  
where $\bm{R}_{\perp}$ ranges over the 2D square lattice projection of the Ti simple-cubic lattice, and $\bm{k}$ is a crystal momentum in the corresponding 2D Brillouin-zone. 

With this notation, the non-interacting Hamiltonian can then be written 
as ${\cal H}_{0} = {\cal T}^{\parallel} + {\cal T}^{\perp}$, where 
the \emph{inter}-layer hopping term is 
\begin{equation}
{\cal T}^{\parallel} = \displaystyle{\sum_{\alpha \sigma}}\displaystyle{\sum_{\bm{k}R_zR_z^{\prime}}}  c^{\dagger}_{\bm{k} R_z \alpha \sigma} \, h^{\parallel}_{\alpha}(R_z-R^{\prime}_z) \, c_{\bm{k} R^{\prime}_z \alpha \sigma},
\end{equation}  
with 
\begin{equation}\label{eq:hoppingZ}
h^{\parallel}_{\alpha}(R_z-R^{\prime}_z)  = \left\{
\begin{array}{l}
{\displaystyle   -t^{\prime}\,\delta_{R_z-R_z^{\prime},\pm a \hat{\bm z}}\quad {\rm for} \,\,\quad \alpha=xy}\vspace{0.2 cm}\\
{\displaystyle  -t\,\delta_{R_z-R_z^{\prime},\pm a \hat{\bm z}} \,\, \quad {\rm for} \,\,\quad \alpha=xz}\vspace{0.2 cm}\\
{\displaystyle  -t\,\delta_{R_z-R_z^{\prime},\pm a \hat{\bm z}}\,\,\quad {\rm for} \,\,\quad \alpha=yz},
\end{array}
\right.~
\end{equation}
and the \emph{intra}-layer hopping term is 
\begin{equation}\label{eq:TperpA}
{\cal T}^{\perp} = \displaystyle{\sum_{\alpha \sigma}} \displaystyle{\sum_{\bm{k} \, R_z}}  c^{\dagger}_{\bm{k} R_z \alpha \sigma} \, h^{\perp}_{\alpha}(\bm{k})\, c_{\bm{k} R_z \alpha \sigma}
\end{equation}  
with 
\begin{equation}\label{eq:hoppingPerp}
h^{\perp}_{\alpha}(\bm{k}) = \left\{
\begin{array}{l}
{\displaystyle   \frac{\hbar^2 {\bm k}^2}{2 m_{\rm L}} -4t \qquad \qquad \qquad \quad \; \; \; \; {\rm for} \qquad \alpha=xy}\vspace{0.2 cm}\\
{\displaystyle   \frac{\hbar^2 k^2_x}{2 m_{\rm L}} +\frac{\hbar^2 k^2_y}{2 m_{\rm H}} -2(t+t^{\prime}) \qquad  {\rm for} \qquad \alpha=xz}\vspace{0.2 cm}\\
{\displaystyle \frac{\hbar^2 k^2_x}{2 m_{\rm H}} +\frac{\hbar^2 k^2_y}{2 m_{\rm L}} -2(t+t^{\prime}) \qquad {\rm for} \qquad \alpha=yz}
\end{array}
\right.~.
\end{equation}

In using Eq.~(\ref{eq:hoppingPerp}) we are assuming that the number of electrons per Ti atom is much less than one,
a condition that is violated only in very narrow quantum well systems,\cite{stemmer_prb_2012,balents_prb_2013} so that 
only states near the $\Gamma$ point of the 2D Brillouin zone are occupied. 
The energy dispersion of the conduction bands can be described using effective masses, which we obtain by expanding
about $\Gamma$ to leading order in $\bm{k}$.  
Eq.~(\ref{eq:hoppingPerp}) shows that the $xz$ and $yz$ bands have one heavy mass, $m_{\rm H}$, and one light mass, $m_{\rm L}$, while the $xy$ band is isotropic and
has a light mass $m_{\rm L}$ in both the $x$ and $y$ directions. 
In terms of the original hopping amplitudes, $m_{\rm H} = \hbar^2/(2 t' a^2)$ and $m_{\rm L} = \hbar^2/(2 t a^2)$. 

In heterojunction systems, several $xy$, $yz$, and $xz$ type subbands are expected to be occupied even at moderate electron 
densities.~\cite{guru_prb_2012,millis_prb_2013,Popovic_prl_2008} However, since $\gtrsim 75\%$ of the 
electron density is contained in the lowest $xy$, $yz$ and $xz$ subbands~\cite{guru_prb_2012}, we address the case in which only 
one subband of each orbital type is retained. 
This truncation is sufficiently realistic to account for the most interesting 
properties of heterojunction 2DEGs.  The assumption can easily be relaxed when there is an 
interest in quantitatively describing the properties of particular 2DEG systems that have more occupied subbands. 
With this assumption Eq.~(\ref{eq:TperpA}) can be written as 
\begin{equation}\label{eq:Tperp}
{\cal T}^{\perp} = \displaystyle{\sum_{\bm{k} \alpha \sigma}} c^{\dagger}_{\bm{k} \alpha \sigma} \, h^{\perp}_{\alpha}(\bm{k})\, c_{\bm{k} \alpha \sigma}~,
\end{equation}  
where $\alpha$ is an orbital label. 
Eq.~(\ref{eq:hoppingPerp}) gives the bare band dispersions for the three bands retained in our model, which are plotted in 
Figure~\ref{fig:One}.  According to recent tight-binding fits~\cite{guru_prb_2012} to Shubnikov-de Haas 
measurements~\cite{Allen_prb_2013} of bulk n-type SrTiO$_3$, $t = 236~{\rm meV}$, and $t' = 35~{\rm meV}$
so that 
\begin{equation}\label{eq:heavymass}
\frac{m_{\rm H}}{m} = \frac{{\rm Ry}}{t'}\left(\frac{a_{\rm B}}{a}\right)^2 \sim 7
\end{equation}
and
\begin{equation}\label{eq:lightmass}
\frac{m_{\rm L}}{m} = \frac{{\rm Ry}}{t}\left(\frac{a_{\rm B}}{a}\right)^2 \sim 1~,
\end{equation}
where $m$ is the bare electron mass, $a_{\rm B}$ is the Bohr radius, and ${\rm Ry}$ is the Rydberg energy unit.  

The isolated-layer bare-electron Hamiltonian in Eq.~(\ref{eq:TperpA}) 
has several important qualitative features that are schematically depicted in Fig.~\ref{fig:One}. 
First, the $xz$ and $yz$ bands have elliptical Fermi surfaces, while the $xy$ band is isotropic. 
Second, the $xz$ and $yz$ band edge is higher in energy than the $xy$ band edge by $2(t-t') \sim 402$ meV
when carriers are confined to a single atomic layer.  
This offset will decrease as the electrons spread out in the confinement direction, but the 
quantum confinement energy will always be lower for $xy$ oribitals.  
Because the $xy$ band electrons have a much weaker hopping amplitude in the
$\hat{\bm z}$-direction ($t^{\prime}$) than the $xz$ and $yz$ electrons ($t$), they are 
more strongly confined to interfaces in heterojunction systems. 
We note that it is possible~\cite{Tolsma_PRB_2016} to crudely account for the difference in interfacial confinement by 
introducing a parameter $d$, which is defined as the difference between $xz$, $yz$ subbands and the $xy$ subband average $\hat{\bm z}$-direction 
positions (see Fig.~\ref{fig:One}), however in the current article we do not make this simplifying approximation.   

Next we add electron-electron interactions to our Hamiltonian. When the electron density per Ti atom is much smaller than one, the Fermi surface occupies a small fraction of the Brillouin zone and the probability of two electrons simultaneously occupying the same Ti site is very small. In this limit, including only the Hubbard part of the full electron-electron interaction misses the most important Coulomb interactions.~\cite{Tolsma_PRB_2016}
Because of its long range, the typical Coulomb interaction energy of an individual electron vanishes only as two-dimensional density $n^{1/2}$, in contrast to the $\propto n$ scaling of the Hubbard model. 
We therefore include in our model calculations the full Coulomb interaction
\begin{equation}
{\cal V} = \frac{1}{2}\displaystyle{\sum_{i \neq j}}\frac{e^2/\kappa}{\sqrt{\left(\bm{r}_{\perp,i}-\bm{r}_{\perp,j} \right)^2 + \left(z_i-z_j\right)^2}}~.
\end{equation}  
Here, $\kappa$ is an effective dielectric constant. 
For the typical electron-electron interaction transition energies, the relevant dielectric constant does not include
the difficult-to-model soft-phonon contribution~\cite{Comment}, but depends on the dielectric environment on both sides of the 
relevant heterojunction or surface.  In order to take advantage of translational symmetry in the $x$-$y$ plane, it will be convenient to 
Fourier transform the Coulomb interaction with respect to the in-plane electron position operators ${\bm r}_{\perp,i}$:
\begin{equation}\label{eq:CoulombQ}
{\cal V} = \frac{1}{2 A}\displaystyle{\sum_{i \neq j}}\displaystyle{\sum_{\bm q}}\, v_{q}(z_i-z_j)\,e^{i\bm{q}\cdot(\bm{r}_{\perp,i} - \bm{r}_{\perp,j} )}
\end{equation}  
where $v_{q}(x)=2 \pi e^2\exp(-q \vert x \vert)/\left( \kappa q\right)$.  We then separate the 
${\bm q}=0$ term in Eq.~(\ref{eq:CoulombQ}), which can be regularized~\cite{Giuliani_and_Vignale,guru_prb_2012,millis_prb_2013}
by combining it with a remote neutralizing background.  
This procedure leads, up to an irrelevant constant, to 
the following Hamiltonian contribution:  
\begin{equation}\label{eq:Hartree}
{\cal V}^{\parallel} = -\frac{1}{2 A}\displaystyle{\sum_{i \neq j}}\frac{2 \pi e^2}{\kappa}\vert z_i -z_j \vert~,
\end{equation} 
which we refer to below as the Hartree contribution. 
Because this term in the Hamiltonian depends only on a small number of macroscopic observables,
namely the number of electrons in each layer, its contribution to the ground-state energy is given exactly by the 
corresponding classical mean-field energy, as discussed in Section~\ref{Subsect:Two_B}.  
The remaining portion of the electron-electron interaction,
\begin{equation}\label{eq:CoulombPerp}
{\cal V}^{\perp} = \frac{1}{2 A}\displaystyle{\sum_{i \neq j}}\displaystyle{\sum_{{\bm q}\neq {\bm 0}}}\,v_{q}(z_i-z_j)\,e^{i\bm{q}\cdot(\bm{r}_{\perp,i}-\bm{r}_{\perp,j} )}~,
\end{equation}  
gives an energy contribution that is dependent on electronic exchange and correlation effects.   
\section{Variational Theory for the ground-state energy}
\label{Sect:Two}
In this Section we develop a variational approximation for the ground-state energy that aims to treat interfacial confinement on
an equal footing with electrostatic and XC interaction energies. 
Our strategy is to first assign confinement wavefunctions to 
each $t_{2g}$ subband, to then use a two-dimensional random-phase (GW) approximation to account for exchange and 
correlation with a fixed confinement-wavefunction constraint, and finally to optimize the confinement wavefunctions 
by minimizing the total energy including electrostatic energies and confinement-direction hopping.
The orbital-dependent subband wavefunctions can be expanded in the form
\begin{equation}\label{eq:wavefunction}
\vert \psi_{\alpha\sigma} \rangle = \sum^{N_l}_{R_z=1} \eta^{\alpha\sigma}_{R_z} \,  \vert R_z  \, \alpha \, \sigma \rangle~,
\end{equation}
where $\vert R_z \, \alpha \, \sigma \rangle $ is a single-particle confinement direction lattice state.  The 
single-particle Hilbert space is constructed as a direct product of 
the $\vert R_z \, \alpha \, \sigma \rangle$ space and two-dimensional momentum space, $\vert \psi_{{\bm k}\alpha \sigma} \rangle  = \vert{\bm k} \rangle \otimes \vert \psi_{\alpha \sigma} \rangle$.   In Sect.~\ref{Subsect:Two_A} we derive an expression for the contribution to the ground-state energy 
from the single-particle and interaction terms that are sensitive to planar 
spatial correlations: ${\cal H}^{\perp} = {\cal T}^{\perp} + {\cal V}^{\perp} $ (see below.) 
Because the number of electrons in each orbital is a good quantum number, the energy can be expressed as a function of
the subband wavefunction spinors, ${\bm \eta}^{\alpha\sigma}$, and 
the electron density associated with each spin-resolved orbital, $n_{\alpha \sigma}$.
In Sect.~\ref{Subsect:Two_B} we derive an equation for the wavefunctions attached to each spin-resolved band
that is based on a total-energy minimization principle.  
Finally, in Sect~\ref{Subsect:Two_C} we describe the self-consistent solution of this 
equation.  Using this procedure we can determine the 
distribution of density amongst the spin-resolved bands, allowing for the possibility of spontaneous 
orbital and spin order.  

\subsection{Two-Dimensional Electron Gas Exchange and Correlation}
\label{Subsect:Two_A}

In this Section we will apply the random phase approximation (RPA)~\cite{Giuliani_and_Vignale} to a 2DEG 
Hamiltonian 
constructed by fixing confinement-direction subband wavefunctions: 
\begin{equation}\label{eq:Hperp}
\begin{array}{l}
{\displaystyle {\cal H}^{\perp} = \displaystyle{\sum_{\alpha \sigma}} c^{\dagger}_{{\bm k}\alpha \sigma}h^{\perp}_{\alpha}(\bm{k}) c_{{\bm k}\alpha \sigma} }\vspace{0.1 cm}\\
{\displaystyle + \; \; \frac{1}{2 A} \; \sum_{{\bm q} \neq 0} \sum_{\mathclap{\substack{ {\bm k} {\bm k}^{\prime} \\ \alpha \alpha^{\prime} \\ \sigma \sigma^{\prime }}}} V_{\alpha \sigma, \alpha^{\prime} \sigma^{\prime}}\left( q \right) \; c^{\dagger}_{{\bm k}+{\bm q} \alpha \sigma}  c^{\dagger}_{{\bm k}^{\prime}-{\bm q} \alpha^{\prime} \sigma^{\prime}} c_{{\bm k}^{\prime} \alpha^{\prime} \sigma^{\prime}} c_{{\bm k} \alpha \sigma} }~.
\end{array}
\end{equation}  
${\cal H}^{\perp}$ accounts fully for lateral hopping and for the component of Coulomb interactions sensitive to the electron coordinates perpendicular to the confinement direction. It is simply the second-quantization version of Eq.~(\ref{eq:CoulombPerp}) combined with Eq.~(\ref{eq:Tperp}). The dispersion of the bands retained in our model is given in Eq.~(\ref{eq:hoppingPerp}) and plotted in Fig.~\ref{fig:One}. The orbital- and spin-dependent effective interaction in Eq.~(\ref{eq:Hperp}) is constructed by attaching the appropriate subband wavefunction to each orbital and averaging over the confinement direction
\begin{equation}\label{eq:V2DEG}
V_{\alpha \sigma, \alpha^{\prime} \sigma^{\prime}} \left( q \right) = 
\sum_{R_{z1},R_{z2}} \,  |\eta^{\alpha \sigma}_{R_{z1}}|^2  \frac{2 \pi e^2 e^{-q|R_{z1}-R_{z2}|}}{\kappa q}  \,  |\eta^{\alpha^{\prime} \sigma^{\prime}}_{R_{z2}}|^2~.
\end{equation} 
In the numerical calculations for the $t_{2g}$ 2DEG presented in Sect.~\ref{Sect:Three} we limit our consideration to wavefunctions that
depend only on orbital, {\it i.e.} ${\bm \eta}^{\alpha \sigma} \rightarrow {\bm \eta}^{\alpha}$, allowing
the spin indices $(\sigma,\sigma^{\prime})$ 
in Eq.~(\ref{eq:V2DEG}) can be dropped. The ground-state energy of the 2DEG Hamiltonian in Eq.~(\ref{eq:Hperp}) has kinetic energy $E_{\rm K}^{\perp}$, exchange energy $E^{\perp}_{\rm X}$, and 
correlation energy $E^{\perp}_{\rm C}$ contributions that depend on a set of densities that specify the occupation 
of each spin-resolved band, $\left\{ n_{\alpha \sigma} \right\}$. The exchange and correlation energies also depend on the set of
transverse subband wavefunctions through the expansion coefficients, $\left\{ {\bm \eta}^{\alpha \sigma} \right\}$ defined in Eq.~(\ref{eq:wavefunction}). Here and throughout the remainder of this Article we use the convention 
\begin{equation}
\left\{x_{\alpha \sigma}\right\} \equiv \left\{x_{xy\,\uparrow} , x_{xy \,\downarrow} , x_{yz \,\uparrow} , x_{yz \,\downarrow} , x_{xz \,\uparrow} , x_{xz \,\downarrow}\right\}
\end{equation}
to express sets of variables that depend on band $\alpha$ and spin $\sigma$ quantum numbers. 

The kinetic energy contribution to the ground-state energy of Eq.~(\ref{eq:Hperp}) is simply the sum of contributions from each occupied spin-resolved band, {\it i.e.}
\begin{equation}
E^{\perp}_{\rm K} = \sum_{\alpha \sigma}E^{\alpha \sigma}_{\rm K}~.
\end{equation}
For each spin-resolved isotropic $xy$ band 
\begin{equation}
E^{\alpha \sigma}_{\rm K}=\frac{\pi \hbar^2 A n^2_{\alpha \sigma}}{m_{\rm L}}~,
\end{equation}
and for each spin-resolved elliptical $xz$ and $yz$ band
\begin{equation}
E^{\alpha \sigma}_{\rm K}=\frac{\pi \hbar^2 A n^2_{\alpha \sigma}}{m_{\rm DOS}}~,
\end{equation}
where $A$ is the 2D sample area and we have defined the density-of-states mass $m_{\rm DOS} = \sqrt{m_{\rm L} m_{\rm H}}$. The offset between the $xz$ and $yz$ band-edge energy and the $xy$ band-edge energy is accounted for in the analysis of ${\cal H}^{\parallel}$ presented in Sect.~\ref{Subsect:Two_B}.

The exchange energy is given by first-order perturbation theory.~\cite{Giuliani_and_Vignale} Since the Coulomb interaction vertices are diagonal in band index, the total exchange energy is given by the sum of independent contributions from each occupied band
\begin{equation}\label{eq:exchange_components}
E^{\perp}_{\rm X} = \displaystyle{\sum_{\alpha \sigma}} E^{\alpha \sigma}_{\rm X}~.
\end{equation}
The exchange energy of electrons in band $\alpha$ and with spin $\sigma$ is given by
\begin{equation} \label{eq:exchange_definition}
E^{\alpha \sigma}_{\rm X} = -\frac{1}{2A}\sum_{\bm k}n^{({\rm F})}_{\alpha \sigma}({\bm k})\sum_{q \neq 0} 
V_{\alpha \sigma, \alpha \sigma}\left( q \right) n^{({\rm F})}_{\alpha \sigma}\left({\bm k}+{\bm q}\right)
\end{equation} 
where $n^{({\rm F})}_{\alpha \sigma}\left({\bm k}\right)$ is the usual Fermi-Dirac distribution function. At zero temperature the integral over ${\bm k}$ can be performed analytically and the remaining two-dimensional integral over ${\bm q}$ can be evaluated numerically:
\begin{widetext}
\begin{equation}
\begin{array}{l}\label{eq:exchange_general}
{\displaystyle E^{\alpha \sigma}_{\rm X} = -\frac{A}{8 \pi^2}\int_0^{\infty} q \, dq \int_0^{2 \pi} d\phi \left[ \sum_{R_{z1},R_{z2}} \,  |\eta^{\alpha \sigma}_{R_{z1}}|^2  \frac{2 \pi e^2 e^{-q F \left(\zeta, \phi \right) |R_{z1}-R_{z2}|}}{\kappa q F \left(\zeta, \phi \right)}  \,  |\eta^{\alpha \sigma}_{R_{z2}}|^2 \right]  \Theta\left( \sqrt{4 \pi n_{\alpha \sigma}} - \frac{q}{2}\right)}\vspace{0.2 cm}\\
{\displaystyle \; \qquad \quad \qquad \qquad \qquad \qquad \qquad \times \; \; \left[ n_{\alpha \sigma} - \frac{q}{4 \pi^2} \sqrt{4 \pi n_{\alpha \sigma} - \left(\frac{q}{2} \right)^2}- \frac{2}{\pi} \arcsin{\left(\frac{q}{4 \sqrt{\pi n_{\alpha \sigma}}} \right)}  \right] }~,
\end{array}
\end{equation} 
\end{widetext} 
where $\Theta\left(x\right)$ is the Heaviside step-function and we have defined 
\begin{equation}
F \left( \zeta,\phi \right) = \sqrt{\zeta^{-1/2}\cos^2{\left( \phi \right)} + \zeta^{1/2}\sin^2{\left( \phi \right)}}~.
\end{equation}
The parameter $\zeta = m_{\rm H}/m_{\rm L}$ in Eq.~(\ref{eq:exchange_general}) describes the degree of band-anisotropy present in the $xz$ and $yz$ bands. Eq.~(\ref{eq:exchange_general}) applies to the isotropic $xy$ bands after setting $\zeta$ to one, in which case the integral over $\phi$ becomes trivial. 

In the numerical calculations presented in Sect.~\ref{Sect:Three} we specifically consider thin-films of SrTiO$_3$. Throughout this Article we define the ``thin-film limit'' quantitatively via the condition $N_la\sqrt{n/g}\ll 1$, where $\sqrt{n/g}$ is an approximation for the 2D Fermi wavevector of each of the $g=6$ spin-resolved bands in our $t_{2g}$ 2DEG model and $N_la$ is the quantum well thickness ({\it i.e.} the maximum distance separating any two electrons in the $z$-direction). Qualitatively this limit describes the case when the typical inter-electron distance which is perpendicular to the confinement direction, $\vert\bm{r}_{\perp,i}-\bm{r}_{\perp,j} \vert$, is much greater than the inter-electron distance parallel to the confinement direction, $\vert z_i - z_j \vert$, and the latter can be neglected. In this limit the Coulomb interaction loses its orbital dependence, and we therefore replace $V_{\alpha \sigma, \alpha \sigma} \left( q \right)$ with $v_q \equiv 2 \pi e^2/\left(\kappa q \right)$. Subsequently, Eq.~(\ref{eq:exchange_general}) for the exchange energy of a single spin-resolved band can be evaluated analytically
\begin{equation}\label{eq:exchange}
 E^{\alpha \sigma}_{\rm X} = - \frac{16 e^2 n_{\alpha \sigma}^{3/2}A}{3 \pi^{3/2} \kappa}\, \zeta^{1/4} \, {\rm K}(1-\zeta)~,
\end{equation}
where ${\rm K}\left(x\right)$ is the complete elliptic integral of the first kind. Eq.~(\ref{eq:exchange}) applies to both the elliptical $xz$ and $yz$ bands and, by setting $\zeta \rightarrow 1$, also to the circular $xy$ band.  

We note that the exchange energy is {\it always negative}, that the exchange interaction acts only
between electrons in the same spin and orbitally resolved band, and that the dependence of exchange energy on density in a single band is superlinear. It follows that exchange favors states in which electrons occupy fewer bands. This is the physical mechanism for the type of ferromagnetism envisioned by Bloch.~\cite{Bloch_1929} In 2DEGs where different bands have different orbital content (e.g.~$xy, yz$ or $xz$), the exchange interaction can lead to orbitally-ordered states, as well as spin-ordered states. We discuss these types of instabilities in detail in Section~\ref{Sect:Three}. It follows from Eq.~(\ref{eq:exchange}) that the exchange energy is weakly reduced by band anisotropy. This property works against exchange-driven spontaneous symmetry breaking in SrTiO$_3$ 2DEGs.

To evaluate the correlation energy we follow the familiar procedure~\cite{Giuliani_and_Vignale} of combining coupling-constant-integration with the fluctuation-dissipation theorem. This method begins by considering the coupling-constant of our Hamiltonian (${\it i.e.}$ the electron's charge) as a tunable parameter: we replace $-e$ with $-e \, \sqrt{\lambda}$ and allow $\lambda$ to vary between zero and one. The ground-state wavefunction $\Psi\left(\lambda\right)$ and total ground-state energy $E\left( \lambda \right)$ both evolve as $\lambda$ is varied. 
At any particular value of $\lambda$, the contribution to the total ground-state energy from the coupling-constant-rescaled Hamiltonian in Eq.~(\ref{eq:Hperp}) is found by taking the expectation value
\begin{equation}
\begin{array}{l}
{\displaystyle E^{\perp}\left(\lambda \right) = \langle \Psi\left(\lambda\right) \vert {\cal H}^{\perp}\left( \lambda \right) \vert \Psi\left(\lambda\right) \rangle}\vspace{0.2 cm}\\
{\displaystyle \quad \quad \quad \; = \langle  {\cal T}^{\perp} \rangle_{\lambda} + \langle {\cal V}^{\perp} \left( \lambda \right) \rangle_{ \lambda}}~.
\end{array}
\end{equation} 
After taking the derivative with respect to $\lambda$ and using the Hellman-Feynman theorem this becomes
\begin{equation}
\frac{dE^{\perp}\left(\lambda \right)}{d \lambda} = \frac{1}{\lambda}\langle {\cal V}^{\perp} \left( \lambda \right) \, \rangle_{\lambda}~.
\end{equation}
The XC energy is then obtained by integrating over the coupling-constant
\begin{equation}\label{eq:excorr_general}
E^{\perp}_{\rm XC} \equiv E^{\perp}\left( 1 \right) - E^{\perp}\left(0\right) = \int_0^1\frac{d\lambda}{\lambda}\langle {\cal V}^{\perp} \left( \lambda \right) \rangle_{\lambda}~.
\end{equation}

The instantaneous structure factor~\cite{Giuliani_and_Vignale} describes correlations between density-fluctuations at wavevector $-{\bm q}$ in band $(\alpha,\sigma)$, and density-fluctuations at wavevector $+{\bm q}$ in band $(\alpha^{\prime},\sigma^{\prime})$, and is defined by
\begin{equation}
 S^{(\lambda)}_{\alpha \sigma , \alpha^{\prime}\sigma^{\prime}}\left({\bm q}\right) = \langle n^{\alpha\sigma}_{-{\bm q}} \, n^{\alpha^{\prime}\sigma^{\prime}}_{\bm q}\rangle_{\lambda}/N~,
\end{equation}
where $n^{\alpha\sigma}_{-{\bm q}} = \sum_{\bm k}c^{\dagger}_{{\bm k}+{\bm q} \alpha \sigma}c_{{\bm k} \alpha \sigma}$ and $N$ is the total number of electrons. Eq.~(\ref{eq:excorr_general}) can then be written
\begin{equation}
\begin{array}{l}
{\displaystyle E^{\perp}_{\rm XC} =\frac{1}{2 A} \int_0^1\frac{d\lambda}{\lambda}  \; \sum_{{\bm q} \neq 0} \sum_{\mathclap{\substack{\alpha \alpha^{\prime} \\ \sigma \sigma^{\prime }}}} \, V^{(\lambda)}_{\alpha \sigma, \alpha^{\prime} \sigma^{\prime}} \left( q \right)}\vspace{0.2 cm}\\
{\displaystyle \; \; \qquad \times \; \; \left[N S^{(\lambda)}_{\alpha \sigma , \alpha^{\prime}\sigma^{\prime}}\left({\bm q}\right)  - \delta_{\alpha\alpha^{\prime}}\delta_{\sigma\sigma^{\prime}} N_{\alpha \sigma}\right] }~,
\end{array}
\end{equation}  
where $N_{\alpha \sigma}$ is the number of electrons in the spin-resolved band labelled by $(\alpha,\sigma)$, and $V^{(\lambda)}_{\alpha \sigma, \alpha^{\prime} \sigma^{\prime}} $ is given by Eq.~(\ref{eq:V2DEG}) after replacing $-e$ with $-e\,\sqrt{\lambda}$. Next we make use of the fluctuation-dissipation theorem~\cite{Giuliani_and_Vignale} relating the instantaneous structure factor to the density-response function
\begin{equation}\label{eq:fluctuation_dissipation}
S^{(\lambda)}_{\alpha \sigma , \alpha^{\prime}\sigma^{\prime}}\left({\bm q}\right) = -\frac{2 \hbar }{n \pi}\int_0^{\infty}\, \chi^{(\lambda)}_{\alpha \sigma , \alpha^{\prime}\sigma^{\prime}}({\bm q}, i \omega)\,d\omega~.
\end{equation}
The density-response function, $\chi^{(\lambda)}_{\alpha \sigma , \alpha^{\prime}\sigma^{\prime}}({\bm q}, i \omega)$, is the proportionality factor between the change in electron density at $-{\bm q}$ in the band labelled by $(\alpha,\sigma)$, when an arbitrarily weak perturbation couples to the density at ${\bm q}$ in the band labelled by $(\alpha^{\prime},\sigma^{\prime})$.~\cite{Giuliani_and_Vignale} We have analytically continued the integration path to the complex frequency axis in Eq.~(\ref{eq:fluctuation_dissipation}) to avoid plasmon poles. We can now write the XC energy as
\begin{equation}
\begin{array}{l}\label{eq:excorr_CC}
{\displaystyle E^{\perp}_{\rm XC} =\frac{1}{2 A} \int_0^1\frac{d\lambda}{\lambda}  \; \sum_{{\bm q} \neq 0} \sum_{\mathclap{\substack{\alpha \alpha^{\prime} \\ \sigma \sigma^{\prime }}}} \, V^{(\lambda)}_{\alpha \sigma, \alpha^{\prime} \sigma^{\prime}} \left( q \right)}\vspace{0.2 cm}\\
{\displaystyle \; \times \; \left[ -\frac{2 \hbar A}{\pi}\int_0^{\infty} \chi^{(\lambda)}_{\alpha \sigma , \alpha^{\prime}\sigma^{\prime}}({\bm q},i \omega)\,d\omega  - \delta_{\alpha\alpha^{\prime}}\delta_{\sigma\sigma^{\prime}} N_{\alpha \sigma}\right] }~.
\end{array}
\end{equation}  

Since the exchange energy is equivalent to first-order perturbation theory in homogeneous electron gases, it is clear that we can obtain an alternative expression for the exchange energy by replacing the full density-response function, $\chi^{(\lambda)}_{\alpha \sigma , \alpha^{\prime}\sigma^{\prime}}({\bm q},i \omega)$, with the non-interacting density-response function, $\chi^{(0)}_{\alpha \sigma , \alpha^{\prime}\sigma^{\prime}}({\bm q},i \omega)$. Note that the non-interacting density-response function is diagonal in the combined spin-band index $(\alpha,\sigma)$, {\it i.e.} $\chi^{(0)}_{\alpha \sigma , \alpha^{\prime}\sigma^{\prime}}({\bm q},i \omega) = \delta_{\alpha\alpha^{\prime}}\delta_{\sigma\sigma^{\prime}}\chi^{(0)}_{\alpha \sigma}({\bm q},i \omega) $. We can subtract the coupling-constant-integration expression for the exchange energy from Eq.~(\ref{eq:excorr_CC}) to obtain the correlation energy
\begin{equation}
\begin{array}{l}\label{eq:corr_CC_one}
{\displaystyle E^{\perp}_{\rm C} =\frac{- \hbar}{\pi}\int_0^{\infty}d\omega \int_0^1\frac{d\lambda}{\lambda}  \; \sum_{{\bm q} \neq 0} \sum_{\mathclap{\substack{\alpha \alpha^{\prime} \\ \sigma \sigma^{\prime }}}} \, V^{(\lambda)}_{\alpha \sigma , \alpha^{\prime} \sigma^{\prime}} \left( q \right)}\vspace{0.2 cm}\\
{\displaystyle \; \qquad \times \; \left[  \chi^{(\lambda)}_{\alpha \sigma , \alpha^{\prime}\sigma^{\prime}}({\bm q},i \omega)  -  \chi^{(0)}_{\alpha \sigma , \alpha^{\prime}\sigma^{\prime}}({\bm q},i \omega)\right] }~.
\end{array}
\end{equation}  
It is, in fact, necessary to remove the exchange-energy contribution from the coupling-constant formulation of the XC energy in order to obtain a convergent frequency integral. 

The density-response functions appearing in Eq.~(\ref{eq:corr_CC_one}) form a matrix, ${\bm \chi}^{(\lambda)}({\bm q},\omega)$, with rows labelled by $(\alpha,\sigma)$ and columns labelled by $(\alpha^{\prime},\sigma^{\prime})$.  In the RPA, the density-response matrix is related to the diagonal matrix of non-interacting density-response functions ${\bm \chi}^{(0)}({\bm q},\omega) \equiv {\rm diag}(\chi^{(0)}_{\alpha_1 \uparrow}({\bm q},\omega),\chi^{(0)}_{\alpha_1 \downarrow}({\bm q},\omega),\chi^{(0)}_{\alpha_2 \uparrow}({\bm q},\omega),\chi^{(0)}_{\alpha_2 \downarrow}({\bm q},\omega),\ldots)$ via
\begin{equation}\label{eq:matrixform}
\big[ {\bm \chi}^{(\lambda)}({\bm q},i\omega)\big]^{-1} = \big[{\bm \chi}^{(0)}({\bm q},i\omega)\big]^{-1} - {\bm V}^{(\lambda)}({\bm q})~,
\end{equation}
where the elements of the matrix ${\bm V}^{(\lambda)}({\bm q})$ are given by Eq.~(\ref{eq:V2DEG}) after replacing $-e$ with $-e \, \sqrt{\lambda}$. Analytic expressions for ${\rm Re}\, \chi^{(0)}_{xy \sigma}({\bm q},i\omega)$ and ${\rm Im}\, \chi^{(0)}_{xy \sigma}({\bm q},i\omega)$ for the isotropic 2D band case can be found, for example, in Ref.~\onlinecite{Giuliani_and_Vignale}. The subband wavefunctions, ${\bm \eta}^{\alpha \sigma}$, do not change these functions from the formulas for a perfectly two-dimensional system. The analytic expressions for the elliptical-band response functions $\chi^{(0)}_{xz \sigma}({\bm q},i\omega)$ [$\chi^{(0)}_{yz \sigma}({\bm q},i\omega)$] can be easily identified by rescaling momenta, $k_x \to k_x \sqrt{m_{\rm L} /m_{\rm DOS}}$ and $k_y \to k_y \sqrt{m_{\rm H} / m_{\rm DOS}}$ [$k_x \to k_x \sqrt{m_{\rm H} / m_{\rm DOS}}$ and $k_y \to k_y\sqrt{m_{\rm L}/m_{\rm DOS}}$] in Eq.~(\ref{eq:hoppingPerp}). Since this rescaling maps the elliptical band onto a circular band, the elliptical band density-response functions $\chi^{(0)}_{xz \sigma}({\bm q},i\omega)$ and $\chi^{(0)}_{yz \sigma}({\bm q},i\omega)$ can be written in terms of $\chi^{(0)}_{xy \sigma}({\bm q},i\omega)$:
\begin{equation}\label{eq:polarizationxz}
\chi^{(0)}_{xz \sigma}({\bm q},i\omega) =  \left. \chi^{(0)}_{xy \sigma}(q',i\omega; m_{\rm DOS}) \right|_{q' \to \sqrt{q^2_x \zeta^{1/2} + q^2_y \zeta^{-1/2}}}
\end{equation}
and
\begin{equation}\label{eq:polarizationyz}
\chi^{(0)}_{yz \sigma}({\bm q},i\omega) =  \left. \chi^{(0)}_{xy \sigma}(q',i\omega; m_{\rm DOS}) \right|_{q' \to \sqrt{q^2_x \zeta^{-1/2} + q^2_y \zeta^{1/2}}}
\end{equation}
where we have again defined $\zeta = m_{\rm H}/m_{\rm L}$. Eq.~(\ref{eq:corr_CC_one}) and Eq.~(\ref{eq:matrixform}) combine to give a general expression for the RPA correlation energy which can be applied to 2DEGs with an arbitrary number of subbands. When the number of bands is sufficiently small, Eq.~(\ref{eq:matrixform}) can be inverted analytically and the integral over the coupling-constant $\lambda$ can be performed by hand. Furthermore, if each band is isotropic then the non-interacting density-response functions depend only on $\vert {\bm q}\vert$ and the integral over wavevector orientation angle is trivial. 

In applying the coupling-constant integration algorithm to the $t_{2g}$ model for SrTiO$_3$, we assume that all electrons in the elliptical $xz$ and $yz$ bands can be described by a single transverse wavefunction, $\bm{\eta}^{\rm E}$, while all electrons in the circular $xy$ band can be described by a second transverse wavefunction, $\bm{\eta}^{\rm C}$. This approximation is motivated by Eq.~(\ref{eq:hoppingZ}), which says that electrons in the elliptical bands hop between layers in the $\hat{\bm z}$-direction with amplitude $t$, while electrons in the circular band hop with amplitude $t^{\prime}$. There are then only three distinct interactions contained in Eq.~(\ref{eq:V2DEG}), {\it i.e.} $V_{{\rm C}, {\rm C}}  = V_{xy \sigma, xy \sigma^{\prime}}$, $V_{{\rm E}, {\rm E}} = V_{xz \sigma, xz \sigma^{\prime}}=V_{yz \sigma, yz \sigma^{\prime}} =V_{xz \sigma, yz \sigma^{\prime}} =V_{yz \sigma, xz \sigma^{\prime}} $, and  $V_{{\rm C}, {\rm E}} = V_{{\rm E}, {\rm C}}  = V_{xy \sigma, xz \sigma^{\prime}} =V_{xy \sigma, yz \sigma^{\prime}}= V_{xz \sigma, xy \sigma^{\prime}} =V_{yz \sigma, xy \sigma^{\prime}}$ where we have omitted the $q$ dependence of the interactions for brevity. Since there are only two independent components, Eq.~(\ref{eq:matrixform}) becomes
\begin{equation}\label{eq:response_simple}
\big[ {\bm \chi}^{(\lambda)}({\bm q},i\omega)\big]^{-1} = 
\left(\begin{array}{ccc}
(\chi^{(0)}_{\rm C})^{-1} & 0\\
0 & (\chi^{(0)}_{\rm E})^{-1} \\
\end{array}
\right) - \left(\begin{array}{ccc}
V^{(\lambda)}_{{\rm C}, {\rm C}} & V^{(\lambda)}_{{\rm C}, {\rm E}} \\
V^{(\lambda)}_{{\rm E}, {\rm C}} & V^{(\lambda)}_{{\rm E}, {\rm E}} \\
\end{array}
\right)~.
\end{equation}
Here, $\chi^{(0)}_{\rm C}({\bm q},i\omega)$ is the density response function of the circular bands,
\begin{equation}
\chi_{\rm C}^{(0)}({\bm q},i\omega)=\sum_{\sigma}\chi^{(0)}_{xy \sigma}(\bm{q},i\omega)~,
\end{equation}
and $\chi_{\rm E}^{(0)}({\bm q},i\omega)$ is the density response function of the elliptical bands,
\begin{equation}
\chi_{\rm E}^{(0)}({\bm q},i\omega)=\sum_{\sigma}\sum_{\alpha = xz, yz}\chi^{(0)}_{\alpha \sigma}(\bm{q},i\omega)~.
\end{equation}

Finally, as in the calculation of the exchange energy presented above, we consider the thin-film limit in which the Coulomb-interaction entering the RPA correlation energy loses its orbital dependence, and we can replace $V_{\alpha \sigma, \alpha^{\prime} \sigma^{\prime}} \left( q \right)$ with $v_q$. Eq.~(\ref{eq:response_simple}) simplifies to a single-component expression
\begin{equation}
\frac{1}{\chi^{(\lambda)}({\bm q},\omega)} = \frac{1}{\chi^{(0)}({\bm q},\omega)} - v^{(\lambda)}_q
\end{equation}
where we define $\chi^{(0)}({\bm q},\omega) = \sum_{\alpha \sigma} \chi_{\alpha \sigma}^{(0)}({\bm q},\omega)$. In this same limit Eq.~(\ref{eq:corr_CC_one}) reduces to 
\begin{equation}
\begin{array}{l}\label{eq:corr_CC_two}
{\displaystyle E^{\perp}_{\rm C} =\frac{- \hbar}{\pi}\int_0^{\infty}d\omega \int_0^1\frac{d\lambda}{\lambda}  \; \sum_{{\bm q} \neq 0}  \, v_q}\vspace{0.2 cm}\\
{\displaystyle \; \qquad \times \; \left[  \chi^{(\lambda)}({\bm q},i \omega)  -  \chi^{(0)}({\bm q},i \omega)\right] }~,
\end{array}
\end{equation}  
which after carrying out the integration over the coupling constant gives
\begin{equation}
\begin{array}{l}\label{eq:corr_CC_three}
{\displaystyle E^{\perp}_{\rm C} =\frac{- \hbar}{\pi}\int_0^{\infty}d\omega \; \sum_{{\bm q} \neq 0}  }\vspace{0.2 cm}\\
{\displaystyle \; \qquad \times \; \left[ v_q\, \chi^{(0)}({\bm q},i \omega)  + \ln\left(1-v_q\,\chi^{(0)}({\bm q},i \omega)  \right)\right] }~.
\end{array}
\end{equation}  

 As explained in more detail in Appendix~\ref{Sect:AppendixA}, we find that the (always negative) correlation energy per electron is decreased in magnitude by increasing the band anisotropy. Furthermore, the correlation energy is largest in magnitude when the electron density is distributed equally amongst all of the spin-resolved bands. This property implies that the correlation energy works against the type of spontaneous symmetry breaking favored by exchange.  We discuss this competition in Section~\ref{Sect:Three}, but first must account for the contribution to the ground-state energy from the $q=0$ electron-electron interaction term ({\it i.e.} the Hartree potential), the electrostatic external potential, and from hopping between layers. 

In this section we have obtained expressions for the contribution to the total ground-state energy from terms in the $t_{2g}$ 2DEG Hamiltonian which are sensitive to electron positions perpendicular to the confinement direction
\begin{equation}\label{eq:energy_perp_total}
E^{\perp} = E^{\perp}_{\rm K} + E^{\perp}_{\rm X} + E^{\perp}_{\rm C}~.
\end{equation}
In the next section we will give expressions for the total ground-state energy of the $t_{2g}$ 2DEG Hamiltonian by combining Eq.~(\ref{eq:energy_perp_total}) with the expectation value of the terms which are sensitive to electron positions parallel to the confinement direction. Before we do this, we pause to note that neither our thin-film approximation, $N_la\sqrt{n/g}  \ll 1$, nor our restriction to only two distinct transverse wavefunctions, $(\bm{\eta}^{\rm C}, \bm{\eta}^{\rm E})$, are critical to the applicability of the variational theory.
\subsection{Total Energy}
\label{Subsect:Two_B}
Next we calculate the contribution to the total ground-state energy from terms in the $t_{2g}$ 2DEG Hamiltonian 
that are sensitive to the electron position in the direction parallel to the confinement 
direction, $ {\cal H}^{\parallel} = {\cal T}^{\parallel} + {\cal V}^{\parallel} + {\cal V}^{\parallel}_{\rm ext} $. This is obtained by taking the expectation value of ${\cal H}^{\parallel}$ in the many-body state defined by
\begin{equation}\label{eq:MB_wavefunction}
 \langle \Psi_{\left\{n_{\alpha \sigma},{\bm \eta}^{\alpha \sigma} \right\}} \vert {\cal H}^{\perp} \vert \Psi_{\left\{n_{\alpha \sigma} , {\bm \eta}^{\alpha \sigma} \right\}}\rangle = E^{\perp}~.
\end{equation}
Although $E^{\perp}$ was not calculated in the previous section by appealing to an explicit many-body wavefunction, the appropriate N-electron wavefunction can always be written as a linear combination of Slater determinants of single-particle states $\vert \psi_{{\bm k}\alpha \sigma} \rangle$:
\begin{equation}\label{eq:MB_wavefunction_explicit}
\vert \Psi_{\left\{n_{\alpha \sigma},{\bm \eta}^{\alpha \sigma} \right\}} \rangle = \sum_{i}c_i \Big\{ \sum_{\rm P} \, (-1)^{\rm P} \, \hat{{\rm P}}\big[\prod^{\rm N}_{j=1} \vert \psi_{{\bm k}_j \alpha_j \sigma_j} \rangle \big] \Big\}~,
\end{equation}
where $\hat{{\rm P}}$ is the permutation operator, ${\rm P}$ is the number of permutations in each term, and we are summing over all permutations of the single-particle state labels amongst the $N$ electrons. Here, $c_i$ is a complex coefficient and the index $i$ runs over all Slater determinants with a fixed number of electrons in each spin-resolved band and fixed transverse wavefuctions; we choose the sets $\left\{ n_{\alpha\sigma}\right\}$ and $\left\{{\bm \eta}^{\alpha \sigma} \right\}$ as good quantum numbers of our many-body variational wavefunction $\vert \Psi_{\left\{n_{\alpha \sigma} , {\bm \eta}^{\alpha \sigma} \right\}} \rangle$.  

The first contribution to $E^{\parallel}$ is from interlayer hopping described by Eq.~(\ref{eq:hoppingZ}). 
Taking the expectation value of ${\cal T}^{\parallel} $ we find that 
\begin{equation}
E^{\parallel}_{\rm K}  = A\displaystyle{\sum_{\alpha \sigma}} n_{\alpha \sigma} \left( {\bm \eta}^{\alpha \sigma \dagger}\cdot{\bm t}^{\alpha}\cdot{\bm \eta}^{\alpha \sigma}\right)~,
\end{equation}
where ${\bm t}^{xz}={\bm t}^{yz}=-t\,\delta_{R_z,R_z\pm a} + 2(t-t^{\prime})\,\delta_{R_z,R_z^{\prime}}$ and ${\bm t}^{xy}=-t^{\prime}\,\delta_{R_z,R_z\pm a} $ are $N_l \times N_l$ matrices in layer index. The diagonal contribution $2(t-t^{\prime})$ to ${\bm t}^{xz}$ and ${\bm t}^{yz}$ accounts for the energy offset between the band edge of the elliptical $xz$ and $yz$ bands and the circular $xy$ band when confined to a single TiO$_2$ layer (see Fig.~\ref{fig:One}).  This expression is independent of the correlations among transverse 
degrees of freedom, approximated above using the RPA.
Similarly, the Hartree energy can be obtained by taking the expectation value of the portion 
of the Coulomb interaction sensitive to electron positions along the confinement direction, ${\cal V}^{\parallel}$:
\begin{equation}
\begin{array}{c}\label{eq:Hartree_energy}
{\displaystyle E^{\parallel}_{\rm H}  = A\displaystyle{\sum_{\alpha \sigma}} n_{\alpha \sigma} \left( {\bm \eta}^{\alpha \sigma \dagger}\cdot{\bm V}_{\rm H}\cdot{\bm \eta}^{\alpha \sigma}\right)}\vspace{0.2 cm}\\
{\displaystyle \quad \quad \quad \; \; \, \; \; = - A  \frac{ \pi e^2}{\kappa}  \sum_{R_z,R_z^{\prime}}  \, n_{R_z} \, \vert R_z-R_z^{\prime} \vert \, n_{R_z^{\prime}} }~,
\end{array}
\end{equation}
where $n_{R_z}$ is the total electron density in layer $R_z$. The matrix ${\bm V}_{\rm H}$ entering Eq.~(\ref{eq:Hartree_energy}) is diagonal in TiO$_2$ layer index, with elements given by 
\begin{equation}
\begin{array}{l}
{\displaystyle \big[ {\bm V }_{\rm H} \big]_{R_z,R_z^{\prime}}= -\delta_{R_z,R_z^{\prime}}\left(\frac{\pi e^2}{\kappa}\right) }\vspace{0.1 cm}\\
{\displaystyle \; \; \; \; \quad \times   \; \; \sum_{R_z^{\prime \prime}} \vert R_z-R_z^{\prime\prime} \vert \, \displaystyle{\sum_{\alpha \sigma}} n_{\alpha \sigma} \vert \eta_{R_z^{\prime\prime}}^{\alpha \sigma} \vert^2  }~.
\end{array}
\end{equation}  
For definiteness, we will assume in this article
the common case in which 
there is only one electrostatic gate (which we place at $z=0$) with total charge $+e\, n = +e \sum_{\alpha \sigma} n_{\alpha \sigma} $ (see Figure~\ref{fig:One}). The electronic energy from the external gate's confinement potential is then
\begin{equation}
E_{\rm ext}^{\parallel}=A\displaystyle{\sum_{\alpha \sigma}} n_{\alpha \sigma} \left( {\bm \eta}^{\alpha \sigma \dagger}\cdot{\bm V}_{\rm ext}\cdot{\bm \eta}^{\alpha \sigma}\right) ~,
\end{equation}
where ${\bm V}_{\rm ext}$ is also a diagonal matrix in TiO$_2$ layer index, with matrix elements given by
\begin{equation}
\big[{\bm V}_{\rm ext}\big]_{R_z,R_z^{\prime}} = \left( \frac{2 \pi e^2 n R_z}{\kappa}\right)\delta_{R_z, R_z^{\prime}}~.
\end{equation}

Combining terms, we write the contribution from the parallel part of the Hamiltonian to the total ground-state energy as
\begin{equation}\label{eq:energy_parallel_total}
 E^{\parallel} = E^{\parallel}_{\rm K} + E^{\parallel}_{\rm H} + E_{\rm ext}^{\parallel}~.
\end{equation}  
Notice that since each term in ${\cal H}^{\perp}$ depends only on a small number of macroscopic observables ({\it e.g.} the density in each layer and/or the density in each spin-resolved band),
which are completely determined from the good quantum numbers of our variational wavefunction Eq.~(\ref{eq:MB_wavefunction_explicit}), each terms contribution to the ground-state energy is given exactly by the corresponding classical mean-field energy. 

When $E^{\parallel}$ is combined with the contributions to the ground-state energy from the perpendicular part of the Hamiltonian, we obtain the total ground-state energy as a functional of the density in each spin-resolved band and the transverse subband wavefunction spinors:
\begin{equation}\label{eq:energy_total_final}
 E\big[\left\{n_{\alpha \sigma} , {\bm \eta}^{\alpha \sigma} \right\} \big] = E^{\parallel} + E^{\perp}~.
\end{equation}  
\subsection{Minimizing the total energy}
\label{Subsect:Two_C}
To solve for the ground-state values of $\left\{n_{\alpha \sigma }\right\}$ and $\left\{\bm{\eta}^{\alpha \sigma }\right\}$, we minimize the energy functional in Eq.~(\ref{eq:energy_total_final}) subject to the constraint equations
\begin{equation}
\left\{
\begin{array}{l}\label{eq:constraints}
{\displaystyle  \sum_{\alpha \sigma} n_{\alpha \sigma} = n }\vspace{0.1 cm}\\
{\displaystyle \displaystyle{\sum_{R_z}}\vert \eta^{\alpha \sigma}_{R_z}\vert^2 = 1}
\end{array}
\right.~.
\end{equation}
The second equation is applied for each $\alpha$ and $\sigma$. For these constraints we introduce Lagrange multipliers, $\mu$ and $\left\{\nu_{\alpha \sigma}\right\}$, respectively. 
Minimization of the total energy functional with respect to each particular $n_{\alpha \sigma}$ yields an equation
\begin{equation}\label{eq:minimization_one}
 \mu^{\perp}_{\alpha \sigma} + {\bm \eta}^{\alpha \sigma \dagger}\cdot \big({\bm t}^{\alpha}+{\bm V}_{\rm H}+{\bm V}_{\rm ext} \big)\cdot{\bm \eta}^{\alpha \sigma} = \mu~,
\end{equation}
 where we have defined $\mu^{\perp}_{\alpha \sigma} = A^{-1}\,dE^{\perp}/dn_{\alpha \sigma}$. In obtaining the results presented in Sect.~\ref{Sect:Three}, we have evaluated $\mu^{\perp}_{\alpha \sigma}$ numerically by combining the above expressions for $E^{\perp}$ with finite difference formulas. Minimization of the total energy functional with respect to each component of the spinor ${\bm \eta}^{\alpha \sigma \dagger}$ yields an eigenvalue equation
\begin{equation}\label{eq:minimization_two}
\big({\bm X} + {\bm Y}^{\alpha \sigma} + {\bm t}^{\alpha \sigma}+{\bm V}_{\rm H}+{\bm V}_{\rm ext} \big) \cdot{\bm \eta}^{\alpha \sigma} = \left(\frac{\nu_{\alpha \sigma}}{A \, n_{\alpha \sigma}}\right){\bm \eta}^{\alpha \sigma} 
\end{equation}
for each value of $\alpha$ and $\sigma$. Here, the diagonal matrices ${\bm X}$ and ${\bm Y}^{\alpha \sigma}$ are found from differentiating the correlation energy and exchange energy, respectively, and their matrix elements are given by 
\begin{widetext}
\begin{equation}
\begin{array}{l}\label{eq:Xmatrix}
{\displaystyle \qquad \qquad \qquad \qquad \big[ {\bm X} \big]_{R_z,R^{\prime}_z} = \delta_{R_z,R^{\prime}_z} \, \left(\frac{-\hbar}{\pi}\right) \int_0^{\infty}d\omega \, \int_0^1\frac{d\lambda}{\lambda}  \; \sum_{{\bm q} \neq 0} \; \sum_{\mathclap{\substack{\alpha \alpha^{\prime} \\ \sigma \sigma^{\prime }}}} \, }\vspace{0.2 cm}\\
{\displaystyle \times \left(M_{\alpha \sigma, \alpha^{\prime} \sigma^{\prime}}\left( R_z,q \right)  \left\{\chi^{(\lambda)}_{\alpha \sigma , \alpha^{\prime}\sigma^{\prime}}  -  \chi^{(0)}_{\alpha \sigma , \alpha^{\prime}\sigma^{\prime}}\right\} + V^{(\lambda)}_{\alpha \sigma, \alpha^{\prime} \sigma^{\prime}}\left(q \right) \; \sum_{\mathclap{\substack{\beta \beta^{\prime} \\ \tau \tau^{\prime }}}} \chi^{(\lambda)}_{\alpha \sigma , \beta \tau} \; M_{\beta \tau, \beta^{\prime} \tau^{\prime}}\left( R_z,q \right) \; \chi^{(\lambda)}_{\beta^{\prime} \tau^{\prime} , \alpha^{\prime} \sigma^{\prime}} \right) }
\end{array}
\end{equation}
and
\begin{equation}
\begin{array}{l}\label{eq:Ymatrix}
{\displaystyle \big[ {\bm Y}^{\alpha \sigma} \big]_{R_z,R^{\prime }_z} = \delta_{R_z,R^{\prime }_z} \, \left(\frac{-A}{8 \pi^2}\right) \int_0^{\infty} \, q \, dq \; \int_0^{2 \pi}\, d\phi  \;\, M_{\alpha \sigma, \alpha \sigma}\left( R_z,q \,F\left[\zeta,\phi \right] \right)\;\, \Theta\left(\sqrt{4 \pi n_{\alpha \sigma}} -\frac{q}{2} \right)  }\vspace{0.2 cm}\\
{\displaystyle \qquad \qquad \qquad \qquad \qquad \quad \times \quad  \left[ n_{\alpha \sigma} - \frac{q}{4 \pi^2} \sqrt{4 \pi n_{\alpha \sigma} - \left(\frac{q}{2} \right)^2}- \frac{2}{\pi} \arcsin{\left(\frac{q}{4 \sqrt{\pi n_{\alpha \sigma}}} \right)}  \right]}~,
\end{array}
\end{equation}
where we have defined
\begin{equation}
M_{\alpha \sigma, \alpha^{\prime} \sigma^{\prime}}\left( R_z,q \right) = \sum_{R^{\prime \prime}_z} \,  \frac{2 \pi e^2 e^{-q|R_{z}-R^{\prime \prime}_{z}|}}{\kappa q}  \,  \left(   |\eta^{\alpha \sigma}_{R^{\prime \prime}_{z}}|^2 + |\eta^{\alpha^{\prime} \sigma^{\prime}}_{R^{\prime \prime}_{z}}|^2 \right)~.
\end{equation}
\end{widetext}
In Eq.~(\ref{eq:Xmatrix}) we have ommitted the $({\bm q},i \omega)$ dependence of the density response functions for brevity. Eqs.~(\ref{eq:constraints})-(\ref{eq:minimization_two}) provide enough independent linear relationships to solve solve 
self-consistently for the various densities, $\left\{n_{\alpha \sigma }\right\}$, and transverse wavefunction spinors $ \left\{\bm{\eta}^{\alpha \sigma }\right\}$, introduced in our model.

As already mentioned above, the numerical results we present in Sect.~\ref{Sect:Three} have been 
obtained  for thin films of SrTiO$_3$, which we identify by the condition $N_l a \sqrt{n/g} \ll 1$. While this approximation is not essential to our variational approach, by making it we simplify the numerical problem to be solved in this first application of our method in which we study orbital and spin ordering instabilities. Specifically, in the thin-film limit the Coulomb interaction, Eq.~(\ref{eq:V2DEG}), loses its orbital dependence and the exchange and correlation energies are no longer directly dependent on the transverse wavefunction spinors (see Eq.~(\ref{eq:exchange}) and Eq.~(\ref{eq:corr_CC_three})). As a result the matrices ${\bm X}$ and ${\bm Y^{\alpha \sigma}}$ vanish. After also making the approximation that only two distinct transverse wavefunction spinors exist, ${\bm \eta}^{\rm C}$ and ${\bm \eta}^{\rm E}$, the constraints in Eq.~(\ref{eq:constraints}) can be written as
\begin{equation}
\left\{  
\begin{array}{l}\label{eq:constraints_B}
{\displaystyle \sum_{\beta = {\rm E},{\rm C}} n_{\beta} = n }\vspace{0.1 cm}\\
{\displaystyle \displaystyle{\sum_{R_z}}\vert \eta^{\rm C}_{R_z}\vert^2 = 1}\vspace{0.1 cm}\\
{\displaystyle \displaystyle{\sum_{R_z}}\vert \eta^{\rm E}_{R_z}\vert^2 = 1 }
\end{array}
\right.
\end{equation}
and Eq.~(\ref{eq:minimization_one}) and Eq.~(\ref{eq:minimization_two}) become 
\begin{equation}\label{eq:minimization_one_B}
 \mu^{\perp}_{\beta} + {\bm \eta}^{\beta}\cdot \left({\bm t}^{\beta}+{\bm V}_{\rm H}+{\bm V}_{\rm ext} \right)\cdot{\bm \eta}^{\beta} = \mu
\end{equation}
and
\begin{equation}\label{eq:minimization_two_B}
\left({\bm t}^{\beta}+{\bm V}_{\rm H}+{\bm V}_{\rm ext} \right)\cdot{\bm \eta}^{\beta} = \left(\frac{\nu_{\beta}}{A \, n_{\beta}}\right){\bm \eta}^{\beta}~, 
\end{equation}
where $\beta = {\rm E}$ or ${\rm C}$.  Eq.~\ref{eq:minimization_one_B} guarantees that all occupied subbands have the 
same chemical potential, while Eq.~\ref{eq:minimization_two_B}  chooses the subband wavefunction as the eigenvalue of the 
mean-field Hamiltonian.  
Appropriate expressions for matrices ${\bm t}^{\beta}$ and ${\bm V}_{\rm H}$, as well as for $\mu^{\perp}_{\beta}$ follow in an obvious manner from the more general definitions in Sect.~\ref{Subsect:Two_B}.  We note in passing that
although the exchange and correlation energies in this approximation
are no longer directly dependent on the transverse wavefunction spinors, they are still sensitive to confinement effects.
For example, the spinors, ${\bm \eta}^{\rm C}$ and ${\bm \eta}^{\rm E}$ are determined
by solving Eq.~(\ref{eq:minimization_two_B}) and are therefore 
sensitive to the band-offset between the $xz$ and $yz$ bands and the $xy$ bands. 
The band-offset obviously influences the equilibrium values of $n_{\beta}$, which in turn affects the values of the exchange and 
correlation energies defined in Eq.~(\ref{eq:exchange}) and Eq.~(\ref{eq:corr_CC_three}).

We now outline one method for obtaining the self-consistent solution. The first step in the determination of the densities $\left(n_{\rm E},n_{\rm C}\right)$, and wavefunctions $\left(\bm{\eta}^{\rm E},\bm{\eta}^{\rm C}\right)$, is to make an initial choice for them which satisfies the equations of constraint. In step two, use these to calculate ${\bm V}_{\rm H}$, $\mu^{\perp}_{\rm E}$ and $\mu^{\perp}_{\rm C}$. Eq.~(\ref{eq:minimization_one_B}) with $\beta={\rm C}$ can be combined with Eq.~(\ref{eq:minimization_one_B}) with $\beta={\rm E}$ to eliminate $\mu$, and root-finding numerical algorithms can be applied to solve for new values of $n_{\rm E}$ and $n_{\rm C}$. These values are then used to re-evaluate the Hartree potential ${\bm V}_{\rm H}$. Eq.~(\ref{eq:minimization_two_B}) is an eigenvalue problem with a real, symmetric, and tridiagonal matrix. It follows that the eigenvalues are real and the eigenvectors can be chosen to have only real components, {\it i.e.} $\left({\bm \eta}^{\beta}\right)^{\dagger} = \left( {\bm \eta}^{\beta} \right)^{\rm T}$. We solve this eigenvalue problem numerically and select the eigenvectors, $\bm{\eta}^{\rm E}$ and $\bm{\eta}^{\rm C}$, which correspond to the smallest eigenvalues. These eigenvalues corresponds to the band-edge energies of the subbands in our model. We keep the wavefunctions corresponding to the smallest energy eigenvalues because we are only interested in keeping the lowest subband of each orbital type $xy$, $xz$ and $yz$. We next return to the aforementioned ``step two'', and repeat until convergence is obtained. Adaptable numerical algorithms exist in the literature~\cite{LAPACK,Numerical_recipes} to assist in solving the eigenvalue problem, finding spline interpolations for the correlation energy, finding roots, and evaluating multi-dimensional integrals.

After a self-consistent solution has been obtained, the eigenvalues of Eq.~(\ref{eq:minimization_two_B}) give an approximation for the band-edge energies of the bands retained in the $t_{2g}$ 2DEG model. Since our variational theory (just like the Hohenberg-Kohn-Sham formulation of density functional theory) is a ground-state theory, we cannot expect that single-particle energies arising therein will give a perfect description of experimental energies. However, we suspect that they will be an improvement over the LDA because, as discussed in the introduction, our method is sensitive to the electron's orbital symmetry. In Sect.~\ref{Sect:Three} we present numerical calculations using the RPA for the matrix of density-response functions ${\bm \chi}^{(\lambda)}({\bm q},i\omega)$. While being exact at high electron density, the RPA neglects vertex corrections which become more important at very low density. Various methods exist for including vertex-correction physics in the matrix of density-response functions (see Ref.~\onlinecite{Giuliani_and_Vignale} and references therein), and we propose that in future calculations the relative success of these approximations can be gauged by comparing the predicted band-edge energy offsets against those measured directly in angle-resolved photoemission experiments.  

\section{Spin and Orbital Order}
\label{Sect:Three}
\begin{figure}[h!]
\includegraphics[width=0.9\linewidth]{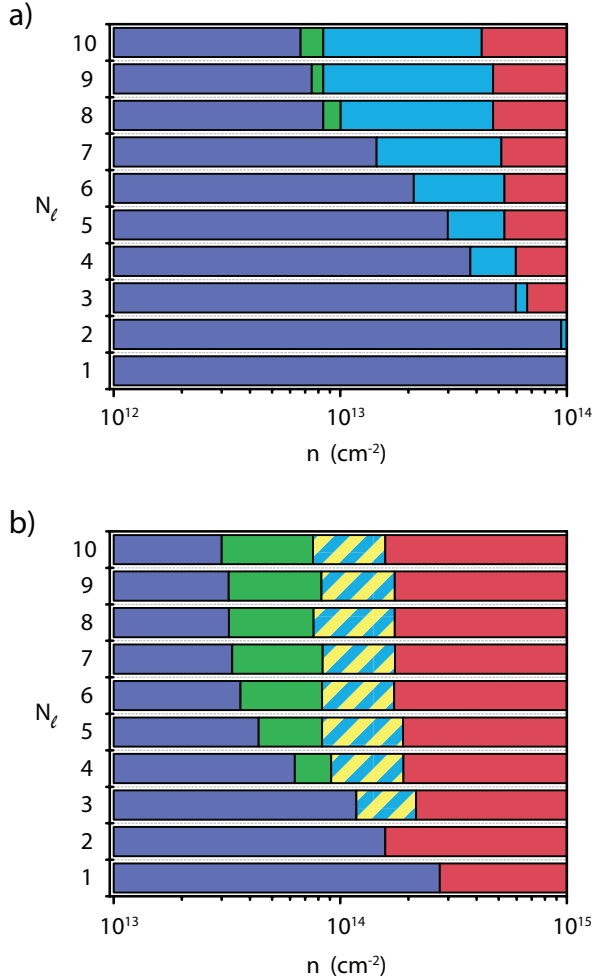}
\caption{(Color online) Phase diagram for the $t_{2g}$ 2DEG model of SrTiO$_3$ plotted for varying total density, $n$, and number of TiO$_2$ layers, $N_l$. In all phases shown there is a finite electron density in the isotropic $xy$ band which is equally distributed between spins. The colors distinguish phases with different distributions of electrons amongst the four spin-resolved elliptical bands ($xz,\uparrow$, $xz,\downarrow$, $yz,\uparrow$, and $yz,\downarrow$). Purple identifies the normal low-density state in which no electrons occupy the elliptical bands. Green identifies a orbital- \emph{and} spin-ordered state in which only one of the spin-resolved elliptical bands contains electrons (e.g.~$xz,\uparrow$). Light blue identifies an orbital-ordered state in which only one elliptical band is occupied (e.g.~$xz,\uparrow$ and $xz,\downarrow$). Yellow identifies a spin-ordered state in which both elliptical bands have electrons but only of one spin species (e.g.~$xz,\uparrow$ and $yz,\uparrow$). Red identifies the normal high-density state in which the density in all four spin-resolved elliptical bands is equal. In panel a) correlations are included within the RPA. In panel b) correlations are neglected and only the exchange energy is included. Although the orbital-ordered state is degenerate with the spin-ordered state in b), correlations prefer an orbital-ordered state and the spin-ordered state is absent in a).\label{fig:Two}}
\end{figure}

In this Section we present the results of our variational calculation for the ground-state energy as a function of $n_{\rm E}$, $n_{\rm C}$, $\bm{\eta}^{\rm E}$, and $\bm{\eta}^{\rm C}$. Specifically, we consider {\it three} types of fully
broken symmetry states characterized by the distribution of electrons in the four spin-resolved elliptical bands: spin-ordered ({\it e.g.} $n_{xz \uparrow}=n_{yz \uparrow} = n_{\rm E}/2$ and $n_{xz \downarrow}=n_{yz \downarrow} = 0$), orbital-ordered ($n_{xz \uparrow}=n_{xz \downarrow} = n_{\rm E}/2$ and $n_{yz \uparrow}=n_{yz \downarrow} = 0$), and orbital- \emph{and} spin-ordered ($n_{xz \uparrow} = n_{\rm E}$ and $n_{xz \downarrow}=n_{yz \uparrow}=n_{yz \downarrow} = 0$). In all of these states we keep the density in the isotropic $xy$ band equally distributed amongst the two spin-species, $n_{xy \uparrow} = n_{xy \downarrow} = n_{\rm C}/2$. The distribution
of electrons in these
 spontaneously broken symmetry states contrasts with the distribution in the normal phase, where all four spin-resolved 
 elliptical bands are either equally occupied (above some critical density which depends on the thickness of SrTiO$_3$), or all empty. Each broken symmetry state has is own set of $n_{\rm E}$, $n_{\rm C}$, $\bm{\eta}^{\rm E}$, and $\bm{\eta}^{\rm C}$, which we vary to find the ground-state energy of that phase. By selecting the state with the lowest energy at each value of total 2DEG density, $n$, and number of TiO$_2$ layers, $N_l$, we have have constructed the phase diagrams shown in Fig.~\ref{fig:Two}.
 
Examining Fig.~\ref{fig:Two}a), we find that spontaneous orbital-order occurs first, and is joined by an accompanying spin-order as the density is decreased. Several striking features appear here. First, the lower critical density for symmetry breaking decreases as the thickness of SrTiO$_3$ increases. Rather than being an interaction effect, this is caused by a reduction in the intrinsic band-edge offset $2(t-t^{\prime})$, see e.g.~Fig.~\ref{fig:One}. Specifically, the elliptical $xz$ and $yz$ bands have a large hopping amplitude parallel to confinement, $t$, and the isotropic $xy$ band has a small hopping amplitude, $t^{\prime}$. Therefore the $xy$ band is more tightly localized to the interface than the $xz$ and $yz$ bands (a result that is quantitatively contained in the differences between $\bm{\eta}^{\rm C}$ and $\bm{\eta}^{\rm E}$, respectively), and the difference in band-edge energies decreases relative to the single-layer value.
This mechanism becomes more pronounced as the sample thickness is increased, leading to a smaller effective band-edge offset at each given total density. Accurately capturing the change in band-edge offset as the total density is varied, and perhaps more importantly as density is allowed to shift between the elliptic bands to the circular bands near an instability, is key to finding the correct phase diagram. In the next Section we elaborate on the role of correlations in selecting the most energetically favorable broken symmetry state.

\subsection{Plasmon-pole approximation: Correlations prefer orbital order}
\label{Subsect:Three_A}

In the isotropic single-band 2DEG, the competition between kinetic and exchange energies qualitatively captures the transition from the normal to the ferromagnetic state.  The role of correlations is to lower
the critical density for symmetry breaking~\cite{Giuliani_and_Vignale}. 
Surprisingly, in our model for thin-film SrTiO$_3$ heterojunctions, we find that the correlation energy plays a 
more qualititave role.  
More specifically, at the exchange-only level the spin-ordered state is energetically degenerate with the orbital-ordered state, as shown by 
the presence of the diagonally-hatched blue and yellow region in Figure~\ref{fig:Two}b). The effective exchange-interaction only acts between electrons in the same spin-resolved band and therefore the ground-state energy is insensitive to whether the $xz,\uparrow$ and $xz,\downarrow$ bands are occupied, as in a orbital-ordered state, or whether the $xz,\uparrow$ and $yz,\uparrow$ bands are occupied, as in a spin-ordered state; the total exchange energy is given in Eq.~(\ref{eq:exchange_components}), and is simply the sum of the exchange energy from each separate spin-resolved band.  Once correlations are included, however, our calculations reveal that the system prefers orbital-order
to spin-order. 
Indeed, the orbitally ordered phase (blue) remains in Fig.~\ref{fig:Two}a), while the spin ordered phase (yellow) is absent. 
\begin{figure}[t]
\includegraphics[width=0.9\linewidth]{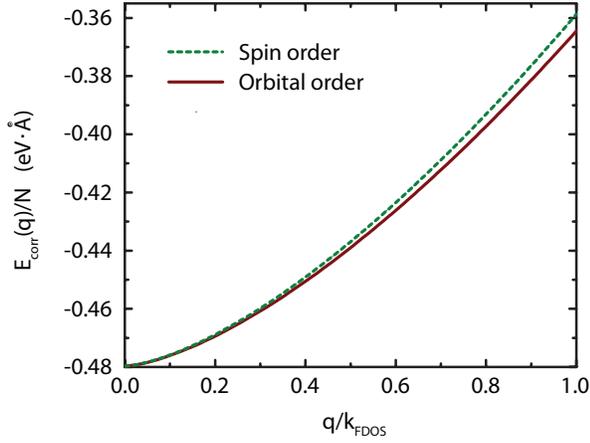}
\caption{(Color online) The contribution to the correlation energy per electron from transitions with wavevector $q$ is plotted for the orbital-ordered (solid brown) and spin-ordered (dashed green) broken symmetry 
states.  (Compare to Eqs.~(\ref{eq:OPenergy}) and Eq.~(\ref{eq:SPenergy}) respectively.)
The correlation energy is calculated within the plasmon-pole approximation and in a simplified model in which the 
circular $xy$ band is absent. We have taken the total density to be $5 \times 10^{12}$cm$^{-2}$. \label{fig:Three}}
\end{figure}

This result can be understood by applying the {\it plasmon-pole approximation} (PPA) to the RPA correlation energy. For simplicity we consider a model without the $xy$ band, and only consider the four remaining bands ({\it i.e.} $\alpha = xz,yz$ and $\sigma=\uparrow,\downarrow$) of our model.
The  plasmon frequency of the spin polarized (SP) and orbitally polarized (OP) states is given by the zeroes of their respective dielectric functions
\begin{equation}\label{eq:dielectricRPA}
\epsilon(\bm{q},\Omega_{i}) = 1-v_q \tilde{\chi}_{i}(\bm{q},\Omega_{i}) = 0~,
\end{equation}
where $i = {\rm OP}$ or ${\rm SP}$, and $\tilde{\chi}_{i}(\bm{q},\omega)$ is the \emph{proper} density response function of the $i$'th broken symmetry state. In the RPA the proper density response function is given by summing the non-interacting density response function of each occupied spin-resolved band. For the orbital-ordered state 
\begin{equation}
\tilde{\chi}_{\rm OP}(\bm{q},\omega) = \chi_{xz,\uparrow}^{(0)}(\bm{q},\omega) + \chi_{xz,\downarrow}^{(0)}(\bm{q},\omega)~,
\end{equation}
and for the spin-polarized state
\begin{equation}
\tilde{\chi}_{\rm SP}(\bm{q},\omega) = \chi_{xz,\uparrow}^{(0)}(\bm{q},\omega) + \chi_{yz,\uparrow}^{(0)}(\bm{q},\omega)~.
\end{equation}
To make analytic progress we will consider plasmons in the long-wavelength limit (i.e.~$q \rightarrow 0$ and $\omega > v_{\rm F} q$). The non-interacting density response function of an isotropic band (e.g~the $xy$ band we are currently neglecting) in this limit is~\cite{Giuliani_and_Vignale}
\begin{equation}
\lim_{q \to 0}\chi_{xy\sigma}^{(0)}(q,\omega) = \frac{n_{xy\sigma} q^2}{m_{\rm L} \, \omega^2}~.
\end{equation}
Expressions for the long-wavelength non-interacting density response functions of the elliptical $xz$ and $yz$ bands can be found from Eq.~(\ref{eq:polarizationxz}) and Eq.~(\ref{eq:polarizationyz}). After inserting the relevant proper density response function into Eq.~(\ref{eq:dielectricRPA}), we find the plasmon frequency of the spin-ordered and orbital-ordered state is
\begin{equation}
\Omega_{\rm SP}=\sqrt{\frac{v_q n}{2 \bar{m}}} q
\end{equation}
and
\begin{equation}
\Omega_{\rm OP}=\sqrt{\frac{v_q n}{m_{\rm DOS}}} \sqrt{\zeta^{\frac{1}{2}}q_x^2 + \zeta^{-\frac{1}{2}}q_y^2}~,
\end{equation}
respectively. Here, $n$ is the total density of the $xz$ and $yz$ bands only since we are not considering the $xy$ band. The density-of-states mass is $m_{\rm DOS}=\sqrt{m_{\rm L}m_{\rm H}}$, the reduced mass is $\bar{m}=m_{\rm L} m_{\rm H}/(m_{\rm L} + m_{\rm H}) $. Interestingly, the SP plasmon becomes isotropic in the long-wavelength limit.

In the PPA, the imaginary part of the density response function is expanded about the plasmon-pole of the system~\cite{Giuliani_and_Vignale}. Using the long-wavelength form of the density response function we obtain 
\begin{equation}
{\rm Im} \, \chi^{\rm PPA}_{i}(q,\omega) = -\frac{\pi \Omega_{i}}{2 v_q}\delta\left(\omega - \Omega_{i}(q)\right).
\end{equation}
The fluctuation-dissipation theorem~\cite{Giuliani_and_Vignale} allows us to then find the instantaneous structure factor
\begin{equation}
S^{\rm PPA}_{i}(\bm{q}) = -\frac{\hbar}{\pi n}\int^{\infty}_0{\rm Im} \, \chi^{\rm PPA}_{i}(\bm{q},\omega) d\omega
\end{equation}
and then the interaction energy per electron in the $i$'th broken symmetry state is given by
\begin{equation}
\frac{E^{i}_{\rm C}}{N} = \frac{1}{8 \pi^2}\int d^2\bm{q} \, v_q\left[S^{\rm PPA}_{i}\left(\bm{q}\right)-1\right].
\end{equation}
Although the long-wavelength PPA static structure factor gives a formally divergent interaction energy, the contribution to the interaction energy at each wavevector $\vert \bm{q} \vert $ has meaning in the long-wavelength limit $\vert \bm{q} \vert \ll k_{\rm F}$~\cite{Pines_and_Nozieres}. We define the correlation energy per electron from transitions at wavevector $\vert \bm{q}\vert$, $E^{i}_{\rm C}\left(q\right)$, by the relationship
\begin{equation}
\frac{E^{i}_{\rm C}}{N} = \int_0^{\infty}\frac{E^{i}_{\rm C}(q)}{N}dq.
\end{equation}
 We find that the contribution to the interaction energy per electron at each wavevector $\vert \bm{q} \vert $ is 
\begin{equation}\label{eq:OPenergy}
\frac{E^{\rm OP}_{\rm C}(q)}{N} = \frac{\hbar q^2}{4 \pi^2 n} \sqrt{\frac{v_q n}{m_{\rm DOS}}} \left[ \zeta^{1/4} \Xi \left(\frac{\zeta -1}{\zeta} \right)  \right]  - \frac{q v_q }{4 \pi}
\end{equation}
in the orbital-ordered state, and in the spin-ordered state is given by
\begin{equation}\label{eq:SPenergy}
\frac{E^{\rm SP}_{\rm C}(q)}{N} = \frac{\hbar q^2}{8 \pi n}\sqrt{\frac{v_q n}{2 \bar{m}}}  - \frac{q v_q }{4 \pi}~,
\end{equation}
where $\Xi(x) $ is the complete elliptic integral of the second kind. The last term on the right hand side in both Eq.~(\ref{eq:OPenergy}) and~(\ref{eq:SPenergy}) is proportional to the classical self-interaction energy of the plasmon density fluctuation at wavelength $2\pi/q$. The first term on the right hand side is the zero-point-energy of each plasmon mode~\cite{Pines_and_Nozieres}.

In Fig.~\ref{fig:Three} we plot Eq.~(\ref{eq:SPenergy}) and Eq.~(\ref{eq:OPenergy}) against $\vert \bm{q} \vert $ in units of an effective density-of-states Fermi wavevector defined by $k_{\rm FDOS} = \sqrt{4 \pi n/g}$, where $g=2$ is the number of occupied spin-resolved bands in either the spin- or orbital-ordered states. The orbitally polarized phase is of lower energy as a result of the smaller zero-point plasmon energy at wavevector $\vert \bm{q} \vert $.
\section{Summary And Discussion}
\label{Sect:Four}
Orbital order is an interesting possibility in $t_{2g}$ electron systems,
even in the itinerant limit of doped $d^{0}$ materials with a small value for the number of 
electrons per metal site.  We refer to the low-density $t_{2g}$
systems as $t_{2g}$ electron gases.  The orbitally ordered $t_{2g}$ electron gas is an electron nematic,~\cite{Fradkin_AnnualReviews_2010} a translationally invariant metallic phase with a spontaneously generated spatial anisotropy. 
In this article we have developed and applied a theory of a two-dimensional $t_{2g}$ electron gases
confined at a surface or interface by an external electric field, which accounts for transverse 
correlations induced by Coulombic electron-electron interactions using a random phase approximation,
and optimizes the subband wavefunctions of the three $t_{2g}$ two-dimensional bands by 
minimizing total energy.  In the present calculation we have assumed that 
the $t_{2g}$ 2DEG has only one subband per flavor, but the formalism allows this approximation 
to be relaxed.  In contrast to a density-functional theory approach, we are 
able in this way to account very specifically for features in the exchange-correlation physics 
that are specific to $t_{2g}$ electron gases, for example 
the difference between the exchange and correlation energies of electrons 
in elliptical {\it vs.} circular bands.

In the ideal isotropic single-band electron gas, the RPA approximation that we employ
for transverse correlations is quantitatively unsuccessful in predicting the Bloch ferromagnetic instability,~\cite{dassarma_prb_2005,attaccalite_prl_2002,tanatar_prb_1989} drastically overestimating the 
density at which it occurs.  However the RPA can be viewed as the leading term in an expansion in 
$N_f^{-1}$, where $N_f$ is the number of fermion bands, and we can expect that it becomes more accurate 
when the number of Fermion species increases.  
This observation suggests that the RPA is a reasonable starting point for treating Coulomb correlations 
in the six band $t_{2g}$ 2DEG model. 

In our calculations we assume that the $t_{2g}$ electrons are confined to an 
interface and also confined in a quantum well with a finite number of metal layers, $N_l$.
Below a critical density that decreases with 
increasing $N_l$ only $xy$ bands are occupied because they have smaller
confinement energies.  We find that orbital order occurs when the 
$yz,xz$ elliptical band density is below a critical value, and that it occurs 
over a broader range of total density for wider quantum wells.  
At very low elliptical band density spin order occurs in additional to orbital order. 

Anisotropic magnetoresistance measurements of LaAlO$_3$/SrTiO$_3$ heterojunctions~\cite{Shalom_prb_2009,Annadi_prb_2013,Ilani_pnas_2013} indicate uniaxial anisotropy which spontaneously appears above a critical density at which the $xz$ and $yz$ bands start to be occupied, and disappears when the $xz$ and $yz$ bands' density becomes large. The magnetic properties of these systems is not perfectly understood to date, but some theoretical analysis based on Hubbard-like model interactions~\cite{Banerjee_NatPhys_2013} and Kondo models~\cite{Ruhman_PRB_2014} has recently appeared. We point out that both the orbitally ordered as well as the orbitally and spin ordered phases we have found would introduce a uniaxial anisotropy in the system, and may explain these measurements.

Several other systems have shown indications of the Bloch-like~\cite{Bloch_1929} instability which drives the transition to orbital and spin order in the $t_{2g}$ electron gas. Experimentally, some evidence for a ferromagnetic instability has been observed in transport studies of very clean low-density Si 2DEGs~\cite{vitkalov_prl_2001,shashkin_prl_2001}. Spin-ordering in these itinerant electron 2DEGs is supported by a large enhancement in the spin-susceptibility as the density is lowered~\cite{pudalov_prl_2002}. Meanwhile, in the multivalley semiconductor AlAs, where the valley degree-of-freedom is closely analogous to our $t_{2g}$ 2DEG's band/orbital degree-of-freedom, large Coulomb-interaction enhancements of the valley susceptibility have been measured~\cite{gunawan_prl_2006}. In addition, exchange-correlation energies are understood to play an important quantitative role in valley occupancy symmetry breaking in Si inversion layers~\cite{Bloss_prb_1980,Bloss_prl_1979}. Lastly, numerical studies of single-band 2DEG systems which include both disorder and the Coulomb interaction have found a magnetic ordered ground-state~\cite{Benenti_PRL_2001,renard_naturecommun_2015}. Despite these studies, Bloch-like broken symmetry states have not been unambiguously observed to date. Our calculations suggest that Bloch-like instability to an orbital-ordered state may be present in thin-film SrTiO$_3$ heterojunctions.

The variational approach described in this Article can be applied to wide 
semiconductor quantum wells in which there is also a subtle interplay between 
correlations and subband wavefunctions, and to Van der Waals heterostructures~\cite{geim_nature_2013}. 
Although multiple orbitals are not often present near the Fermi energy, these layered materials have a weak electronic
hopping amplitude in the confinement direction, and as a result many subbands are occupied even in relatively thin multilayers. 
Additionally, many van der Waals materials have been observed to be strongly interacting, often exhibiting diverse phenomena like charge-density wave instabilities~\cite{xi_natnano_2015}, superconductivity~\cite{Lu_science_2015}, and negative compressibility.~\cite{Larentis_NanoLett_2014} 
\begin{acknowledgments}
JRT thanks Jeremy Levy, Lu Li, Jos\'e Lorenzana and Fengcheng Wu for interesting discussions. 
Work in Austin was supported by the DOE Division of Materials Sciences and Engineering under 
grant DE-FG02-ER45118 and by the Welch foundation under grant TBF1473. MP was supported by Fondazione Istituto Italiano di Tecnologia and the European Union's Horizon 2020 research and innovation programme under Grant Agreement No. 696656 ``GrapheneCore1''.
\end{acknowledgments}

\appendix
\section{The RPA correlation energy dependence on number of bands and band anisotropy}
\label{Sect:AppendixA}
\begin{figure}[t]
\includegraphics[width=0.95\linewidth]{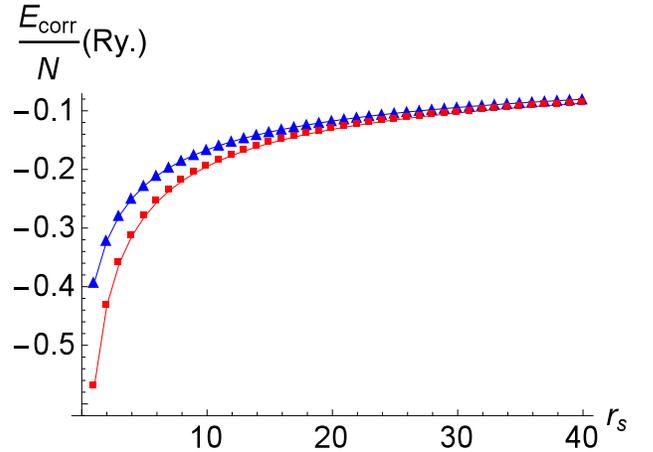}
\caption{(Color online) A plot of the the RPA correlation energy per electron (in units of effective Rydbergs) as the parameter $r_{\rm s}$ is varied from one to forty. Eq.~(\ref{eq:Appendix_1}) defines $r_{\rm s}$ in terms of the total electron density. The blue curve is for a single band isotropic 2DEG with equal populations of both spin species. The red curve is for a 2DEG with two identical isotropic bands with equal spin populations in both bands. Note that the bands are of different orbital content, and therefore interband scattering is negligeable. The correlation energy per electron increases in magnitude with the number of occupied bands. \label{fig:Appendix_one}}
\end{figure}

In this Appendix we explore the dependence of the RPA correlation energy on the number of bands with distinct orbital character, and on the degree of band anisotropy. In Figure~\ref{fig:Appendix_one} we examine how the correlation energy changes by the addition of a second identical isotropic band of effective mass $m^{\rm eff}$. We plot the correlation energy in units of effective Rydberg ${\rm Ry.} = m^{\rm eff} {\rm ry.}/\kappa^2$ where ${\rm ry.} = 13.6$ meV is the bare Rydberg. The parameter $r_{\rm s}$ in this figure is defined in terms of the total 2DEG density, $n$, which is distributed equally amongst either one or two bands, 
\begin{equation}\label{eq:Appendix_1}
r_{\rm s} = \frac{m^{\rm eff}}{a_{\rm B} \kappa \sqrt{\pi n}}~.
\end{equation}
Since the dielectric constant $\kappa$ and the effective mass $m^{\rm eff}$ enter into the definition of both ${\rm Ry.}$ and $r_{\rm s}$, the curves in Figure~\ref{fig:Appendix_one} are independent of the exact numerical value of both $\kappa$ and $m^{\rm eff}$. We find that the RPA correlation energy per electron is larger for the system with two bands. 

\begin{figure}[t]
\includegraphics[width=0.95\linewidth]{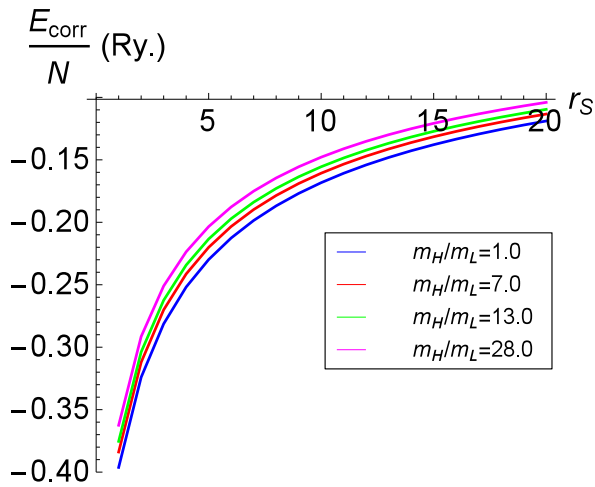}
\caption{(Color online) A plot of the RPA correlation energy per electron (in units of effective Rydberg) for single band 2DEGs of varying band anisotropy defined by the ratio of the mass in the $x$-direction, $m_{\rm H}$, and the mass in the $y$-direction, $m_{\rm L}$. The correlation energy per electron is reduced in magnitude for increasing band anisotropy. \label{fig:Appendix_two}}
\end{figure}

In Figure~\ref{fig:Appendix_two} we examine how the correlation energy of a single 2D band changes as the degree of anisotropy varies. We take the band to have an elliptical Fermi surface and a heavy mass, $m_{\rm H}$, in the $x$-direction and a light mass, $m_{\rm L}$, in the $y$-direction (exactly like the $yz$ band of our model for SrTiO$_3$ in the main text). We plot the correlation energy in units of effective Rydberg ${\rm Ry.} = \sqrt{m_{\rm H}m_{\rm L}} {\rm ry.}/\kappa^2$. The parameter $r_{\rm s}$ in this figure is defined in terms of the total 2DEG density $n$,
\begin{equation}
r_{\rm s} = \frac{\sqrt{m_{\rm H}m_{\rm L}}}{a_{\rm B} \kappa \sqrt{\pi n}}~.
\end{equation}
We find that the correlation energy per electron is reduced as the degree of band anisotropy increases.

\begin{thebibliography}{99}
%
\bibitem{goodenough_pr_1955}
	J. B. Goodenough, 
	\href{http://dx.doi.org/10.1103/PhysRev.100.564}
	{Phys. Rev.~{\bf 100}, 564 (1955)}.
%
\bibitem{wollan_pr_1955}
	E. O. Wollan and W. C. Koehler, 
	\href{http://dx.doi.org/10.1103/PhysRev.100.545}
	{Phys. Rev.~{\bf 100}, 545 (1955)}.
%
\bibitem{cheng_prl_2008}
	J.-G. Cheng, Y. Sui, J.-S. Zhou, J. B. Goodenough and W. Su, 
	\href{http://dx.doi.org/10.1103/PhysRevLett.101.087205}
	{Phys. Rev. Lett.~{\bf 101}, 087205 (2008)}.
%
\bibitem{yan_prl_2004}
	J.-Q. Yan, J.-S. Zhou and J. B. Goodenough, 
	\href{http://dx.doi.org/10.1103/PhysRevLett.93.235901}
	{Phys. Rev. Lett.~{\bf 93}, 235901 (2004)}.
%
\bibitem{ren_prb_2009}
        J.-S. Zhou, J. B. Goodenough, J.-Q. Yan, J.-G. Cheng, K. Matsubayashi, Y. Uwatoko and Y. Ren
	\href{http://dx.doi.org/10.1103/PhysRevB.80.224422}
	{Phys. Rev. B.~{\bf 80}, 224422 (2009)}.
%
\bibitem{mizokawa_prb_1996}
	T. Mizokawa and A. Fujimori, 
	\href{http://dx.doi.org/10.1103/PhysRevB.54.5368}
	{Phys. Rev. B. {\bf 54}, 5368 (1996)}.
%
\bibitem{mochizuki_jpsj_2004}
	M. Mochizuki and M. Imada, 
	\href{http://dx.doi.org/10.1143/JPSJ.73.1833}
	{J. Phys. Soc. Jpn. {\bf 73}, 1833 (2004)}.
%
\bibitem{mannhart_science_1010}
	J. Mannhart and D. G. Schlom, 
	\href{http://dx.doi.org/10.1126/science.1181862}{Science~{\bf 327}, 1607 (2010)}.
%
\bibitem{stemmer_apl_2013}
        A. P. Kajdos, D. G. Ouellette, T. A. Cain, and S. Stemmer, 
        \href{http://dx.doi.org/10.1063/1.4819203}
	{Appl. Phys. Lett.~{\bf 103}, 082120 (2013)}.
%
\bibitem{hwang_apl_2010}
        Y. Kozuka, M. Kim, H. Ohta, Y. Hikita, C. Bell, and H. Y. Hwang, 
        \href{http://dx.doi.org/10.1063/1.3524198}
	{Appl. Phys. Lett.~{\bf 97}, 222115 (2010)}.
%
\bibitem{stemmer_armr_2014}
	S. Stemmer and S. J. Allen, 
	\href{http://dx.doi.org/10.1146/annurev-matsci-070813-113552}
	{Annu. Rev. Mater. Res.~{\bf 44}, 151 (2014)}.
%
\bibitem{levy_armr_2014}
	J. A. Sulpizio, S. Ilani, P. Irvin, and J. Levy, 
	\href{http://dx.doi.org/10.1146/annurev-matsci-070813-113437}
	{Annu. Rev. Mater. Res.~{\bf 44}, 117 (2014)}.
%
\bibitem{Tolsma_PRB_2016}
	J. R. Tolsma, A. Principi, R. Asgari, M. Polini, and A. H. MacDonald, 
	\href{http://dx.doi.org/10.1103/PhysRevB.93.045120}
	{Phys. Rev. B.~{\bf 93}, 045120 (2016)}.
%
\bibitem{Bloch_1929}
	F. Bloch, Z. Physik~{\bf 57}, 545 (1929).
%
\bibitem{young_nature_1999}
	D. P. Young, D. Hall, M. E. Torelli, Z. Fisk, J. L. Sarrao, J. D. Thompson, H.-R. Ott, S. B. Oseroff, 
	R. G. Goodrich, and R. Zysler, \href{http://dx.doi.org/10.1038/17081}{Nature~{\bf 397}, 412 (1999)}.
%
\bibitem{vitkalov_prl_2001}
	S. A. Vitakalov, H. Zheng, K. M. Mertes, M. P. Sarachik, and T. M. Klapwijk, 
	\href{http://dx.doi.org/10.1103/PhysRevLett.87.086401}
	{Phys. Rev. Lett.~{\bf 87}, 086401 (2001)}.
%
\bibitem{shashkin_prl_2001}
        A. A. Shashkin, S. V. Kravchenko, V. T. Dolgopolov, and T. M. Klapwijk, 
        \href{http://dx.doi.org/10.1103/PhysRevLett.87.086801}
	{Phys. Rev. Lett.~{\bf 87}, 086801 (2001)}.
%
\bibitem{teneh_prl_2012}
	N. Teneh, A. Yu. Kuntsevich, V. M. Pudalov, and M. Reznikov, 
	\href{http://dx.doi.org/10.1103/PhysRevLett.109.226403}{Phys. Rev. Lett.~{\bf 109}, 226403 (2012)}.
%
\bibitem{takashina_prb_2013}
	K. Takashina, Y. Niida, V. T. Renard, B. A. Piot, D. S. D. Tregurtha, A. Fujiwara, and Y. Hirayama, 
	\href{http://dx.doi.org/10.1103/PhysRevB.88.201301}{Phys. Rev. B~{\bf 88}, 201301(R) (2013)}.
%
\bibitem{renard_naturecommun_2015}
	V. T. Renard, B. A. Piot, X. Waintal, G. Fleury, D. Cooper, Y. Niida, D. Tregurtha, A. Fujiwara, 
	Y. Hirayama, and K. Takashina, \href{http://dx.doi.org/10.1038/ncomms8230}{Nature Commun.~{\bf 6}, 7230 (2015)}.
%
\bibitem{dassarma_prb_2005}
	Y. Zhang and S. Das Sarma, 
	\href{http://dx.doi.org/10.1103/PhysRevB.72.115317}
	{Phys. Rev. B.~{\bf 72}, 115317 (2005)}.
%
\bibitem{attaccalite_prl_2002}
	C. Attaccalite, S. Moroni, P. Gori-Giorgi and G.B. Bachelet, 
	\href{http://dx.doi.org/10.1103/PhysRevLett.88.256601}
	{Phys. Rev. Lett.~{\bf 88}, 256601 (2002)}.
%
\bibitem{tanatar_prb_1989}
	B. Tanatar and D. Ceperley, 
	\href{http://dx.doi.org/10.1103/PhysRevB.39.5005}
	{Phys. Rev. B.~{\bf 39}, 5005 (1989)}.
%
\bibitem{Hohenberg_Kohn_PhysRev_1964}
	P. Hohenberg and W. Kohn, 
	\href{http://dx.doi.org/10.1103/PhysRev.136.B864}
	{Phys. Rev.~{\bf 136}, B864 (1964)}.
%
\bibitem{Kohn_Sham_PhysRev_1965}
	W. Kohn and L. J. Sham, 
	\href{http://dx.doi.org/10.1103/PhysRev.140.A1133}
	{Phys. Rev.~{\bf 140}, A1133 (1965)}.
%
\bibitem{Langreth_PRB_1980}
	D. C. Langreth and J. P. Perdew, 
	\href{http://dx.doi.org/10.1103/PhysRevB.21.5469}
	{Phys. Rev. B.~{\bf 21}, 5469 (1980)}.
%
\bibitem{Perdew_PRL_1996}
	J. P. Perdew, K. Burke, and M. Ernzerhof,
	\href{http://dx.doi.org/10.1103/PhysRevLett.77.3865}
	{Phys. Rev. Lett.~{\bf 77}, 3865 (1996)}.
%
\bibitem{Giuliani_and_Vignale}
	G. F. Giuliani and G. Vignale, {\it Quantum Theory of the Electron Liquid} 
	(Cambridge University Press, Cambridge, 2005).
%
\bibitem{Jones_RMP_1989}
	R. O. Jones and O. Gunnarsson,
	\href{http://dx.doi.org/10.1103/RevModPhys.61.689}
	{Rev. Mod. Phys.~{\bf 61}, 689 (1989)}.
%
\bibitem{Popovic_prl_2008}
	Z. S. Popovic, S. Satpathy, and R. M. Martin,
	\href{http://dx.doi.org/10.1103/PhysRevLett.101.256801}
	{Phys. Rev. Lett.~{\bf 101}, 256801 (2005)}.
%
\bibitem{guru_prb_2012}
	G. Khalsa and A.H. MacDonald, 
	\href{http://dx.doi.org/10.1103/PhysRevB.86.125121}
	{Phys. Rev. B.~{\bf 86}, 125121 (2012)}.
%
\bibitem{millis_prb_2013}
	S.E. Park and A.J. Millis, 
	\href{http://dx.doi.org/10.1103/PhysRevB.87.205145}
	{Phys. Rev. B.~{\bf 87}, 205145 (2013)}.
%
\bibitem{Ueno_natmater_2008}
	K. Ueno, S. Nakamura, H. Shimotani, A. Ohtomo, N. Kimura, T. Nojima, 
	H. Aoki, Y. Iwasa, and M. Kawasaki, 
	\href{http://dx.doi.org/10.1038/nmat2298}
	{Nature Mater.~{\bf 7}, 855 (2008)}.
%
\bibitem{ilani_natcomm_2014}
	A. Joshua, S. Pecker, J. Ruhman, E. Altman, and S. Ilani, 
	\href{http://dx.doi.org/10.1038/ncomms2116}
	{Nature Commun.~{\bf 3}, 1129 (2012).}
%
\bibitem{stemmer} Santosh Raghavan, S. James Allen, and Susanne Stemmer,
Appl. Phys. Lett. {\bf 103}, 212103  (2013).
%
\bibitem{Goldman_PRL_2011}
        Y. Lee, C. Clement, J. Hellerstedt, J. Kinney, L. Kinnischtzke, X. Leng, 
        S. D. Snyder, and A. M. Goldman,
	\href{http://dx.doi.org/10.1103/PhysRevLett.106.136809}
	{Phys. Rev. Lett.~{\bf 106}, 136809 (2011)}.
%
\bibitem{stemmer_prb_2012}
	P. Moetakef, C. A. Jackson, J. Hwang, L. Balents, 
	S. J. Allen, and S. Stemmer, 
	\href{http://dx.doi.org/10.1103/PhysRevB.86.201102}
	{Phys. Rev. B.~{\bf 86}, 201102(R) (2012)}.
%
\bibitem{balents_prb_2013}
	R. Chen, S. Lee, and L. Balents, 
	\href{http://dx.doi.org/10.1103/PhysRevB.87.161119}
	{Phys. Rev. B.~{\bf 87}, 161119(R) (2013)}.
%
\bibitem{Allen_prb_2013}
        S. J. Allen, B. Jalan, S. Lee, D. G. Ouellette, G. Khalsa, 
        J. Jaroszynski, S. Stemmer, and A. H. MacDonald,
	\href{http://dx.doi.org/10.1103/PhysRevB.88.045114}
	{Phys. Rev. B.~{\bf 88}, 045114 (2013)}.
%
\bibitem{Comment}
	Although the long-wavelength low-temperature static dielectric constant of bulk SrTiO$_3$ 
	is in the tens of thousands [A. S. Barker Jr. and M. Tinkham, 
	\href{http://dx.doi.org/10.1103/PhysRev.125.1527}
	{Phys. Rev.~{\bf 125}, 1527 (1962)}] because of the 
	presence of a soft LO phonon mode near the $\Gamma$ point, 
	the effective dielectric constant which screens electron-electron interactions 
	is expected to be substantially smaller. The electronic transitions which contribute 
	to the ground-state energy in the $t_{2g}$ model are on the scale
	of a few hundred meV, much larger than the soft-phonon mode energy which is closer 
	to a few meV [Y. Yamada and G. Shirane, 
	\href{http://dx.doi.org/10.1143/JPSJ.26.396}{J. Phys. Soc. Jpn.~{\bf 26}, 396 (1969)}]. 
	Screening by this mode is therefore weak. 
	Furthermore, the LO phonon mode is soft only in close vicinity to the $\Gamma$ point. 
	Even static electric fields are ineffectively screened by this mode unless they are constant 
	over distances which greatly exceed a lattice constant. Based on these considerations, 
	we think that the effective dielectric constant to be included in electron-electron interaction 
	calculations is closer to $\kappa \sim 15$, similar to small-gap covalent semiconductors. At liquid helium temperatures the dielectric constant of bulk SrTiO$_3$ 
	is $\kappa \sim 10$ in the range $2-40$ meV and $\kappa \sim 7$ above $40$ meV 
	[R. C. Neville, B. Hoeneisen and C.A. Mead, 
	\href{http://dx.doi.org/10.1063/1.1661463}{J. Appl. Phys.~{\bf 43}, 2124 (1972)}].  
%
\bibitem{Gradshteyn_and_Ryzhik}
	I. S. Gradshteyn and I. M. Ryzhik, {\it Table of Integrals, Series, and Products} 
	(Elsevier Academic Press, 2007).
%
\bibitem{macdonald_prb_1994}
	L. Zheng and A. H. MacDonald, 
	\href{http://dx.doi.org/10.1103/PhysRevB.49.5522}
	{Phys. Rev. B.~{\bf 49}, 5522 (1994)}.
%
\bibitem{LAPACK}
	E. Anderson, Z. Bai, C. Bischof, S. Blackford, J. Demmel, J. Dongarra, J. Du Croz, A. Greenbaum, S. Hammarling, A. McKenney, and D. Sorensen, {\it LAPACK Users' Guide} 
	(Society for Industrial and Applied Mathematics, Philadelphia, 1999).
%
\bibitem{Numerical_recipes}
	W. Press, S. A. Teukolsky, W. T. Vetterling, and B. P. Flannery, {\it Numerical Recipes in C} 
	(Cambridge University Press, Cambridge, 2002).
%
\bibitem{Pines_and_Nozieres} 
	D. Pines and P. Nozi\'eres, {\it The Theory of Quantum Liquids} 
	(W.A. Benjamin, Inc., New York, 1966).
%
\bibitem{Fradkin_AnnualReviews_2010}
	E. Fradkin, S. A. Kivelson, M. J. Lawler, J. P. Eisenstein and A. P. Mackenzie, 
	\href{http://dx.doi.org/10.1146/annurev-conmatphys-070909-103925}
	{Annu. Rev. Condens. Matter Phys.~{\bf 1}, 153 (2010)}.
%
\bibitem{Shalom_prb_2009}
	M. Ben Shalom, C. W. Tai, Y. Lereah, M. Sachs, E. Levy, D. Rakhmilevitch, A. Palevski and Y. Dagan, 
	\href{http://dx.doi.org/10.1103/PhysRevB.80.140403}
	{Phys. Rev. B.~{\bf 80}, 140403(R) (2009)}.
%
\bibitem{Annadi_prb_2013}
	A. Annadi, Z. Huang, K. Gopinadhan, X. Renshaw Wang, A. Srivastava, Z. Q. Liu, H. Harsan Ma, T. P. Sarkar, T. Ventatesan and Ariando, 
	\href{http://dx.doi.org/10.1103/PhysRevB.87.201102}
	{Phys. Rev. B.~{\bf 87}, 201102(R) (2012)}.
%
\bibitem{Ilani_pnas_2013}
	A. Joshua, J. Ruhman, S. Pecker, E. Altman and S. Ilani, 
	\href{http://dx.doi.org/10.1073/pnas.1221453110}
	{Proc. Nat. Acad. Sci. ~{\bf 110}, 9633 (2013)}.
%
\bibitem{Banerjee_NatPhys_2013}
	S. Banerjee, Onur Erten, and M. Randeria,
	\href{http://dx.doi.org/10.1038/NPHYS2702}
	{Nature Phys.~{\bf 9}, 626 (2013)}.
%
\bibitem{Ruhman_PRB_2014}
	J. Ruhman, A. Joshua, S. Ilani, and E. Altman, 
	\href{http://dx.doi.org/10.1103/PhysRevB.87.201102}
	{Phys. Rev. B.~{\bf 90}, 125123 (2014)}.
%
\bibitem{pudalov_prl_2002}
	V. M. Pudalov, M. E. Gershenson, H. Kojima, N. Butch, E. M. Dizhur, 
	G. Brunthaler, A. Prinz, and G. Bauer, 
	\href{http://dx.doi.org/10.1103/PhysRevLett.88.196404}{Phys. Rev. Lett.~{\bf 88}, 196404 (2002)}.
%

\bibitem{gunawan_prl_2006}
	O. Gunawan, Y. P. Shkolnikov, K. Vakili, T. Gokmen, 
	E. P. De Pootere, and M. S. Shayegan, \href{http://dx.doi.org/10.1103/PhysRevLett.97.186404}
	{Phys. Rev. Lett.~{\bf 97}, 186404 (2006)}.
%
\bibitem{Bloss_prb_1980}
        W. L. Bloss, S. C. Ying, and J. J. Quinn,
	\href{http://dx.doi.org/10.1103/PhysRevLett.23.1839}
	{Phys. Rev. B.~{\bf 23}, 1839 (1980)}.
%
\bibitem{Bloss_prl_1979}
        W. L. Bloss, L. J. Sham, and V. Vinter,
	\href{http://dx.doi.org/10.1103/PhysRevLett.43.1529}
	{Phys. Rev. Lett.~{\bf 43}, 1529 (1979)}.
%
\bibitem{Benenti_PRL_2001}
	G. Benenti, G. Caldara, and D. L. Shepelyansky,
	\href{http://dx.doi.org/10.1103/PhysRevLett.86.5333}
	{Phys. Rev. Lett.~{\bf 86}, 5333 (2001)}.
%
\bibitem{geim_nature_2013}
	A. K. Geim, and I. V. Grigorieva, 
	\href{http://dx.doi.org/10.1038/nat12385}
	{Nature~{\bf 499}, 419 (2013)}.
%
\bibitem{xi_natnano_2015}
	X. Xi, L. Zhao, Z. Wang, H. Berger, L. Forro, J. Shan, and K. F. Mak 
	\href{http://dx.doi.org/10.1038/nnano.2015.143}
	{Nature Nanotech.~{\bf 10}, 765 (2015)}.
%
\bibitem{Lu_science_2015}
	J. M. Lu, O. Zheliuk, I. Leemakers, N. F. Q. Yuan, U. Zeitler, 
	K. T. Law, and J. T. Ye, 
	\href{http://dx.doi.org/10.1126/science.aab2277}{Science~{\bf 350}, 1353 (2015)}.
%
\bibitem{Larentis_NanoLett_2014}
        S. Larentis, J. R. Tolsma, B. Fallahazad, D. C. Dillen, K. Kim, A. H. MacDonald, and E. Tutuc, 
        \href{http://dx.doi.org/10.1021/nl500212s}
	{Nano. Lett.~{\bf 4}, 2039 (2010)}.
%
\end{thebibliography}
\end{document}